\documentclass[aps,twocolumn]{revtex4}
\usepackage{epsfig}
\usepackage{amsfonts}

\begin{document}

\title{
Stochastic approach to equilibrium and nonequilibrium thermodynamics}

\author{T\^ania Tom\'e and M\'ario J. de Oliveira}

\affiliation{Instituto de F\'{\i}sica,
Universidade de S\~{a}o Paulo, \\
Caixa Postal 66318\\
05314-970 S\~{a}o Paulo, S\~{a}o Paulo, Brazil}

\date{\today}

\begin{abstract}

We develop the stochastic approach to thermodynamics based on
the stochastic dynamics, which can be discrete (master equation)
continuous (Fokker-Planck equation), 
and on two assumptions concerning entropy.
The first is the definition of entropy itself and
the second, the definition of entropy production rate
which is nonnegative and vanishes in thermodynamic equilibrium.
Based on these assumptions we study interacting systems
with many degrees of freedom in equilibrium or out
of thermodynamic equilibrium, and how the macroscopic laws are
derived from the stochastic dynamics.
These studies include the quasi-equilibrium processes,
the convexity of the equilibrium surface,
the monotonic time behavior of thermodynamic potentials,
including entropy, the bilinear form of the entropy production rate,
the Onsager coefficients and reciprocal relations,
and the nonequilibrium steady states of chemical reactions.


\end{abstract}

\maketitle

%-------------------------------------------------------
\section {Introduction}

The kinetic theory, introduced and developed in the second
half of the nineteenth century by Clausius \cite{clausius1857},
Maxwell \cite{maxwell1860},
and Boltzmann \cite{boltzmann1872,boltzmann1877}
aimed to derive the macroscopic properties of matter,
which include the laws of thermodynamics, from the underlying
microscopic movement, governed by the laws of mechanics. In principle,
this task can be achieved if we assume that the macroscopic laws are
connected to the microscopic laws. Whether the 
connection exists or not we cannot known a priori.
But, considering that it is an experimental fact that
the laws of mechanics, classical or quantum, are obeyed at the microscopic
level by the same particles that constitute a macroscopic body,
which obeys macroscopic laws, it is reasonable to assume that the
connection exists. 
Once we assume this connection, the next task is to perform
the actual derivation of 
macroscopic laws from the microscopic laws of mechanics.
This task was in fact undertaken by the founders
of the kinetic theory and many macroscopic laws  were in fact derived.
This includes the theorem of equipartition of energy,
the Maxwell distribution of velocities \cite{maxwell1860},
the Boltzmann H-theorem \cite{boltzmann1872} and
the Gibbs probability distribution \cite{gibbs1902}.
However, many results cannot be said to have been derived
from pure mechanics alone \cite{dunbar1982,walstad2012}.
A new ingredient was introduced in the course of derivation, namely, 
the stochastic behavior, in most cases in an implicit form.

The derivation from pure mechanics of the results just mentioned
would be accomplished if we could show that the
new ingredient, the stochastic behavior, is a consequence of 
the microscopic mechanical motion, which is deterministic.
At first sight the random behavior seems to be 
in contradiction with a deterministic motion.
However, the results coming from the theory of 
deterministic chaos \cite{schuster1984}
has proven that a deterministic
motion can behave stochastically.
In fact, the possibility of mapping chaotic dynamics
into a stochastic process has already been addressed \cite{nicolis1988}.
The Gibbs probability distribution, for instance, 
is believed to come from the underlying mechanics
through a stochastic behavior,
although there is no known general derivation from pure mechanics.
In some cases, 
the derivation is know \cite{sinai1977}.
In other cases, such as a system of hard spheres,
numerical simulations
of the equations of motion may, for instance, show the
validity of the equipartition of energy or may 
provide the macroscopic properties directly \cite{alder1957}.

The reasoning and examples given above lead us to
presume that the macroscopic properties are obtained
from microscopic mechanics in two major steps:
(1) from the underlying
mechanics to a probabilistic or stochastic approach and 
(2) from this approach to the macroscopic properties.
This is particularly clear
in the case of equilibrium thermodynamic properties,
which are derived from the Gibbs probability distribution,
which in turn comes from the underlying mechanics, a step
not yet fully demonstrated and known as ergodic hypothesis.
The first step will not concern us here.
The second step, which is the purpose of the present paper,
aims to derive the macroscopic
properties, which include equilibrium and nonequilibrium
thermodynamic properties, from a stochastic approach.

The stochastic approach to equilibrium and nonequilibrium
thermodynamics or, in short, stochastic thermodynamics,
which is the second step of our scheme
and the subject of the present paper,
has been adopted by several authors
and become a consistent theory of nonequilibrium thermodynamics
\cite{thomsen1953,klein1954,klein1955,hill1966,mcquarrie1967,schnakenberg1976,
jiuli1984,mou1986,gillespie1992,perez1994,gaveau1997,sekimoto1998,
tome1997,mazur1999,crochik2005,zia2006,
andrieux2006,tome2006,
schmiedl2007,harris2007,zia2007,seifert2008,blythe2008,gaveau2009a,
gaveau2009b,esposito2009,tome2010,broeck2010,tome2012,spinney2012,esposito2012,
zhang2012a,seifert2012,zhang2012b,ge2012,santillan2013,luposchainsky2013,wu2014}.
An important step in this direction occurred when Schnakenberg
\cite{schnakenberg1976}
introduced the stochastic definition of entropy production rate
which has a fundamental role in our approach
in addition to the probabilistic definition of entropy itself,
introduced by Boltzmann \cite{boltzmann1877} and generalized by
Gibbs \cite{gibbs1902}.

Our approach here is based on the adoption of a Markovian
stochastic evolution on a discrete or on a continuous space
and on two assumptions concerning entropy. The first is 
the definition of entropy itself and the second,
the definition of entropy production rate. Based on these
assumptions we will consider systems in equilibrium and out
of thermodynamic equilibrium and how the macroscopic
nonequilibrium laws can be derived from the 
stochastic dynamics, which is the second step mentioned above.
We will treat some fundamental issues that
have barely been considered or that
have not been addressed in the context of stochastic thermodynamics.
This includes several thermodynamic results of systems in 
equilibrium \cite{callen1960,oliveira2013} and out of equilibrium
\cite{donder1927,onsager1931,denbigh1951,
prigogine1955,groot1962,glansdorff1971,nicolis1977}
such as the quasi-equilibrium processes,
the convexity of the equilibrium surface in thermodynamic space,
the monotonic time behavior of thermodynamic potentials,
including entropy, the bilinear form of the entropy production rate,
the Onsager coefficients and reciprocal relations,
and the nonequilibrium steady states of chemical reactions.

The stochastic approach in continuous state space was used
by Einstein \cite{einstein1905}, 
Smoluchowski \cite{smoluchowski1906} and Langevin \cite{langevin1908}
to explain the Brownian motion. 
It was generalized to the case of Brownian particles
subject to an external force by Fokker \cite{fokker1914},
Smoluchowski \cite{smoluchowski1915}, Planck \cite{planck1917},
and Ornstein \cite{ornstein1917}, and the equation governing the
time evolution of the probability distribution became known
as the Fokker-Planck equation. 
Kramers \cite{kramers1940} extended the Fokker-Planck equation
to the case of a massive particle and 
studied the escape of a Brownian particle over a potential barrier
arriving at the Arrhenius factor.

Markovian stochastic dynamics
\cite{kampen1981,gardiner1983,risken1984,tome2015}
has been used in various problems in physics, chemistry and biology,
either in continuous or discrete state space.
In the former case, the evolution of the probability distribution
is governed by a Fokker-Planck equation and in the later by
a master equation.
We mention the  study of chemical reactions
\cite{mcquarrie1967,mou1986,gillespie1992,schmiedl2007,ge2012},
population dynamics and epidemiology 
\cite{bailey1957,bartlett1960,nisbet1982},
and biological systems in general
\cite{hill1966,hill1989,berry2003,andrieux2006,ao2008,
lan2012,zhang2012a,zhang2012b,england2013}.
We wish to mention particularly the stochastic models
with many degrees of freedom such as the so called
stochastic lattice models usually used to describe 
phase transitions and criticality in physics, chemistry and biology
\cite{crochik2005,glauber1963,harris1974,liggett1985,ziff1986,satulovsky1994,
oliveira1992,marro1999,oliveira2003}.

%----------------------------------------------------------
\section{Master equation}

%--------------------------------------
\subsection{Entropy and entropy production}

We assume that the system follows a microscopic stochastic 
dynamics. More precisely, we assume that the system is
described by a continuous time Markovian stochastic process.
Considering a discrete space of states, this assumption
amounts to say that the time evolution equation is set up
once the transition rates are given.
The transition rates
play thus a fundamental role in the present approach and we
may say that a system is considered to be theoretically defined when this 
quantity is given a priori. Given the transition rates,
the probability $P_i(t)$ of state $i$ at time $t$
is obtained by solving the evolution equation,
in this case a master equation, 
\begin{equation}
\frac{d}{dt}P_i(t) = \sum_j \{W_{ij}P_j(t) - W_{ji}P_i(t)\},
\label{21}
\end{equation}
where $W_{ij}$ denotes the transition rate from state $j$ to state $i$.
In this section and the next we will consider transitions
with the following property:
if the rate $W_{ij}$ of the transition $j\to i$ is nonzero then the 
rate $W_{ji}$ of the reverse transition $i\to j$ is also nonzero. 
Later on, in the study of the Fokker-Planck,
we will have the opportunity to treat the case in which the reverse
transition rate may vanish.

As mentioned above, the derivation of the macroscopic properties,
including the laws of thermodynamics, is carried out by the
introduction of two assumptions concerning entropy.
The first is the definition of entropy itself.
The entropy $S$ of a system in equilibrium or out of
equilibrium is taken to be the expression
\begin{equation}
S(t) = -k_B\sum_i P_i(t)\ln P_i(t),
\label{6}
\end{equation}
which is the extension of the equilibrium Boltzmann-Gibbs
entropy to nonequilibrium situations,
where $k_B$ is the Boltzmann constant.

The second assumption concerns the definition of the production of entropy.
This quantity should meet two fundamental properties.
It must be nonnegative and vanish identically in thermodynamic
equilibrium.
Following Schnakenberg \cite{schnakenberg1976},
we assume the following expression for the entropy production rate
\begin{equation}
\Pi(t) = \frac{k_B}2 \sum_{ij} \{W_{ij}P_j(t) - W_{ji}P_i(t)\}
\ln\frac{W_{ij}P_j(t)}{W_{ji}P_i(t)},
\label{9}
\end{equation}
which is clearly nonnegative because each term is of the
form $(x-y)\ln(x/y)$.
This form of entropy production rate has been used by
several authors \cite{jiuli1984,mou1986,crochik2005,tome2006,
zia2006,andrieux2006, schmiedl2007,zia2007,esposito2009,
tome2010,tome2012,esposito2012,tome2015}
within stochastic dynamics and applications.

%--------------------------------------
\subsection{Entropy flux}

Let us consider the time variation of the average
of a state function such as energy, given by
%$U=\langle E_i\rangle$.
\begin{equation}
U(t) = \sum_i E_i P_i(t).
\label{11}
\end{equation}
Using the master equation (\ref{21}) it follows that
\begin{equation}
\frac{dU}{dt} = \Phi_u.
\label{17}
\end{equation}
where 
\begin{equation}
\Phi_u(t) = \sum_{ij} (E_i-E_j) W_{ij} P_j(t),
\label{18}
\end{equation} 
is the total flux of energy from outside to the system.
Equation (\ref{17}) represents the conservation of energy.

Equation of the type (\ref{17}) is valid for any
conserved quantity and this is not the case of entropy.
For instance, in a nonequilibrium stationary state 
the total flux of energy vanishes but not the
total flux of entropy, which is nonzero because entropy is
continuously being produced. The equation for the time
variation of entropy $S$ should be written as \cite{prigogine1955}
\begin{equation}
\frac{dS}{dt} = \Pi - \Phi,
\label{24}
\end{equation}
where $\Phi$ is the flux of entropy from the system
to the outside and $\Pi$ is the entropy production per
unit time, given by equation (\ref{9}).
It is common to write $d_iS/dt$ and $d_eS/dt$ for the entropy
production rate and entropy flux, respectively, but we avoid
this terminology because these quantities are not in fact
time derivatives of any quantity.

Taking the time derivative of equation (\ref{6}) and 
using the master equation (\ref{21}),
we may write the time derivative of entropy as
\begin{equation}
\frac{dS}{dt} = k_B\sum_{ij}
\{W_{ij} P_j(t) - W_{ji}P_i(t)\} \ln P_i(t),
\label{26a}
\end{equation} 
or, in an equivalent form,
\begin{equation}
\frac{dS}{dt} = k_B\sum_{ij}
W_{ij} P_j(t) \ln\frac{P_i(t)}{P_j(t)}.
\label{26}
\end{equation} 
Comparing with (\ref{24}) we see that 
the right hand side of this equation should equal $\Pi-\Phi$.
Using the definition of $\Pi$, given by (\ref{9}),
which we write in the form
\begin{equation}
\Pi(t) = k_B \sum_{ij} W_{ij}P_j(t)
\ln\frac{W_{ij}P_j(t)}{W_{ji}P_i(t)},
\label{9r}
\end{equation}
and comparing with equation (\ref{26})
we get the flux of entropy from the system to outside
\begin{equation}
\Phi(t) = k_B \sum_{ij} W_{ij}P_j(t)\ln\frac{W_{ij}}{W_{ji}},
\label{9f}
\end{equation}
which is equivalent to 
\begin{equation}
\Phi(t) = \frac{k_B}2 \sum_{ij}\{ W_{ij}P_j(t) - W_{ji}P_i(t) \}
\ln\frac{W_{ij}}{W_{ji}}.
\label{9fl}
\end{equation}

The integration of (\ref{24}) in a time interval
will lead us to the Clausius inequality. Indeed,
from equation (\ref{24}) we may write
\begin{equation}
\Delta S = \int \Pi dt - \int \Phi dt.
\end{equation}
If we identify the entropy flux $\Phi$
as the ratio between the heat flux $dQ/dt$ and 
the temperature $T$ of the environment,
then $\int\Phi dt=-\int(dQ/T)$.
But the first integral is nonnegative because $\Pi\geq0$
so that
\begin{equation}
\Delta S \geq \int \frac{dQ}T,
\end{equation}
which is the Clausius inequality \cite{clausius1865}.
In equilibrium, $\Delta S=\int dQ/T$, equality that was used by
Clausius to define entropy.
The difference between $\Delta S$ and the integral $\int dQ/T$,
which is the production of entropy, represents according to
Clausius the ``uncompensated transformation'' \cite{clausius1865}. 
%''uncompensirte (sic) Verwandlung''
%''unkompensierte Verwandlung''

It is common in the recent literature to use another
nomenclature for
the entropy production $\Pi$, the entropy flux $\Phi$
and the time derivative of entropy $dS/dt$.
The quantities that correspond to the
time integral of these three quantities are
called, respectively, the total entropy change, the 
environment entropy change and internal entropy change
\cite{seifert2012,luposchainsky2013}.

%--------------------------------------
\subsection{Thermodynamic equilibrium}

The microscopic definition of
thermodynamic equilibrium, from the static point of view,
is usually characterized in terms of the Gibbs probability distribution.
From the dynamic point of view, the description of equilibrium
by the Gibbs distribution is necessary but
not sufficient. There are examples
\cite{kunsch1984,godreche2011,oliveira2011}
of spin models that 
are described by the Gibbs distribution but are not
in thermodynamic equilibrium in the sense that entropy
is continuously being generated.
From a dynamic point of view, the thermodynamic equilibrium
is characterized by the vanishing of the entropy production rate
and, of course, by a time independent probability distribution.
The vanishing of (\ref{9}) gives
\begin{equation}
W_{ij}P_j = W_{ji}P_i,
\label{8b}
\end{equation}
which is the detailed balance condition, that characterizes
the thermodynamic equilibrium \cite{klein1955},
and equivalent to microscopic reversibility.

In the stationary state, that is,
when the probability $P_i$
is independent of time, the right hand side of (\ref{21})
vanishes, that is,
\begin{equation}
\sum_j \{W_{ij} P_j - W_{ji} P_i \}=0,
\label{27}
\end{equation} 
which we may call global balance. 
The reversibility condition (\ref{8b}) can thus be understood as
detailed balance condition because each term
of the global balance equation vanishes.
Although the global balance is a necessary condition for reversibility
it is not a sufficient condition.

Considering that the equilibrium distribution $P_i^e$
is known, the solution of (\ref{8b}) for the transition rate is 
\begin{equation}
W_{ij} = K_{ij}
\left(\frac{P_i^e}{P_j^e}\right)^{1/2},
\end{equation}
where $K_{ij}$ is symmetric, that is,
$K_{ij}=K_{ji}$.
The transition rates for the various situation in which
the system is found in equilibrium in the stationary state
can now be constructed.
For an isolated system (microcanonical ensemble) the 
equilibrium probability distribution $P_i$ is
a constant whenever the energy function $E_i$ equals 
a given energy, say $U$, and vanishes otherwise. 
Therefore, in this case $W_{ij} = K_{ij}$ when $E_i=E_j$
and vanishes otherwise. In short, $W_{ij} = W_{ji}$.

For a system in contact with a heat reservoir (canonical ensemble)
at temperature $T$, the equilibrium probability distribution
is given by
\begin{equation}
P_i^e = \frac{1}{Z} e^{-\beta E_i},
\label{13}
\end{equation}
where $\beta=1/k_BT$, so that, in this case the transition rate 
fulfils the relation
\begin{equation}
\frac{W_{ij}}{W_{ji}} =  e^{-\beta (E_i-E_j)},
\end{equation}
and is given by
\begin{equation}
W_{ij} = K_{ij} e^{-\beta (E_i-E_j)/2}.
\label{22}
\end{equation}

If, in addition to be in contact with a heat reservoir, the
system is in contact with a reservoir of particles, then
\begin{equation}
P_i^e = \frac{1}{\Xi}e^{-\beta E_i +\beta\mu n_i},
\end{equation}
where $\mu$ is the chemical potential and $n_i$ is the number
of particles. In this case the transition rate fulfils the relation
\begin{equation}
\frac{W_{ij}}{W_{ji}} = 
e^{-\beta (E_i-E_j)+\beta\mu(n_i-n_j)},
\label{23a}
\end{equation}
and is given by
\begin{equation}
W_{ij} = K_{ij} 
e^{-\beta (E_i-E_j)/2+\beta\mu(n_i-n_j)/2},
\label{23}
\end{equation}
where again $K_{ij}=K_{ji}$.

%--------------------------------------
\subsection{The approach to equilibrium}

Let us consider the transient regime of a system
that approaches equilibrium. The time dependent
probability distribution is the solution of the 
master equation (\ref{21})
with transition rates that satisfy the detailed balance
and is appropriate for each type of contact of the system
with the environment.

We treat first the case of microcanonical
distribution, which describes an isolated system. In this
case, as we have seen, $W_{ij}=W_{ji}$
so that the entropy flux (\ref{9f}) vanishes identically, $\Phi=0$.
Therefore,
\begin{equation}
\frac{dS}{dt} = \Pi,
\end{equation}
so that
\begin{equation}
\frac{dS}{dt} \geq 0.
\end{equation}
That is, the entropy of an isolated system is a monotonically
increasing function of time.

Next we consider the canonical distribution which describes
the contact of a system with a heat reservoir.
The transition rate is given by (\ref{22}), which
replaced in the entropy flux (\ref{9f}) gives
\begin{equation}
\Phi = 
-k_B\beta\sum_{ij} \{ W_{ij}P_j(t) - W_{ji}P_i(t) \} E_i.
\end{equation}
Using the master equation (\ref{21}),
the flux of entropy can be written in the form
\begin{equation}
\Phi = -\frac1T\frac{dU}{dt},
\label{35}
\end{equation}
where $U$ is the average of energy, given by (\ref{11}). Equation (\ref{35})
shows that the quantity $\Phi$ is proportional to $dU/dt$.
Notice that (\ref{35}) implies that $\Phi$ vanishes 
in the equilibrium regime ($t\to\infty$) as it should.

Equation (\ref{24}) gives
\begin{equation}
\frac{dU}{dt}-T\frac{dS}{dt} =  - T\Pi.
\end{equation}
If we define the free energy by
$F=U-TS$ and take into account that $T$ is constant,
that is, does not depend on time, we get
\begin{equation}
\frac{dF}{dt} =  - T\Pi,
\end{equation}
so that
\begin{equation}
\frac{dF}{dt} \leq 0.
\label{36}
\end{equation}
That is, the free energy of a system in contact with a heat
reservoir is a monotonically decreasing function of time.
In other terms, the free energy decreases monotonically
to its equilibrium value.

Equation (\ref{36}) is also the
expression of the Boltzmann H-theorem \cite{boltzmann1872}.
Indeed, the Boltzmann $H$ function is defined by
\begin{equation}
H(t) = \sum_i P_i(t)\ln\frac{P_i(t)}{P_i^{\,e}},
\end{equation}
where $P_i^{\,e}$ is the equilibrium canonical distribution
given by equation (\ref{13}). It is straightforward to show that
$F=F_0 +H/\beta$, where $F_0$ does not depend on time.
Therefore, the inequality (\ref{36}) is equivalent to $dH/dt\leq 0$,
which is the Boltzmann H-theorem.

The grand canonical distribution describes the 
contact of the system with a particle reservoir and with a heat
reservoir. The transition rate for this case is given by
(\ref{23}), which replaced in the expression (\ref{9f}) 
and using the master equation (\ref{21})
allows to reach the following expression for the entropy flux 
\begin{equation}
\Phi = - \frac1T\frac{dU}{dt} + \frac{\mu}{T} \frac{dN}{dt},
\end{equation}
where $U$ is the average energy, given by (\ref{11}), and
$N$ is the average number of particles,
\begin{equation}
N(t) = \sum_i n_i P_i(t).
\label{19}
\end{equation}
Taking into account that $dS/dt=\Pi-\Phi$, we get
\begin{equation}
 \frac{dU}{dt} - T\frac{dS}{dt}-  \mu\frac{dN}{dt} = -T\Pi,
\label{28}
\end{equation}
which can be written as 
\begin{equation}
\frac{d\phi}{dt} = -T\Pi,
\end{equation}
where $\phi=U-TS-\mu N$ is the grand thermodynamic potential
and we have taken into account that
$T$ and $\mu$ are constant. Since $\Pi\ge0$ it follows
that $d\phi/dt\leq0$.

Let us integrate equation (\ref{28}) from an initial
time $t=t_0$ to infinity,
\begin{equation}
(U-U_0) - T(S-S_0)-\mu(N-N_0) = -T\int_{t_0}^\infty \Pi dt,
\end{equation}
from which follows the inequality
\begin{equation}
(U-U_0) - T(S-S_0)-\mu(N-N_0) \leq 0,
\label{34}
\end{equation}
because $\Pi\geq0$. Taking into account that,
for large enough times, the system reaches equilibrium
at a temperature $T$ and imposing that at
$t=t_0$ the system was in equilibrium, at a different temperature, 
say $T_0$, we may conclude from the inequality (\ref{34})
that $U$, $S$ and $N$ make up a
convex surface, in accordance with equilibrium thermodynamics.

%--------------------------------------
\subsection{Quasi-equilibrium}

It is common to state the laws of equilibrium thermodynamics
in terms of thermodynamic processes. 
This seems at first sight contradictory because
a process implies a change in the thermodynamic state
and thus a displacement from equilibrium.
To overcome this problem, one introduces
the quasi-static
process, a process which is so slow that the
system may be considered to be in equilibrium.
We will show below that the production of entropy
in this process is negligible so that in fact the
system may be considered to be in equilibrium.
In which sense the production is negligible will be seen below.

Let us consider a system in contact with a heat bath and a particle
reservoir whose temperature and chemical potential,
understood as control parameters,
are slowly varying in time. To describe this situation
we assume a time dependent transition rate $W_{ij}(t)$
of the form (\ref{23}), where $K_{ij}(t)$ may  depend o time,
that is,
\begin{equation}
W_{ij}(t) = K_{ij}(t) 
e^{-\beta (E_i-E_j)/2+\beta\mu(n_i-n_j)/2},
\label{151}
\end{equation}
where $K_{ij}(t)=K_{ji}(t)$, so that
\begin{equation}
\frac{W_{ij}(t)}{W_{ji}(t)} =
\frac{e^{-\beta (E_i-\mu n_i)}}
{e^{-\beta (E_j-\mu n_j)}},
\label{151a}
\end{equation}
where $\beta=1/k_BT$ and $T(t)$ depends on time and
the chemical potential $\mu(t)$ also depends on time.
We assume moreover that $d\beta/dt=\alpha$ 
and $d\mu/dt=\gamma$ are small and are both of the
same order of magnitude.

Replacing equation (\ref{151a}) in expression (\ref{9f})
for the entropy flux and after some straightforward
algebraic manipulation we reach again the result
\begin{equation}
\Phi = - \frac1T\frac{dU}{dt} + \frac{\mu}{T} \frac{dN}{dt}.
\label{161}
\end{equation}
Now $dS/dt=\Pi-\Phi$ so that
\begin{equation}
 \frac{dU}{dt} = T\frac{dS}{dt} + \mu\frac{dN}{dt}-T\Pi.
\label{171}
\end{equation}

%----------------------------
\begin{figure}
\epsfig{file=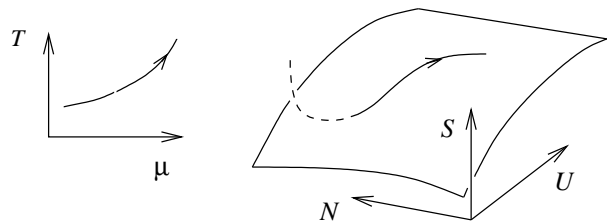,width=8cm}
\caption{A path in the $T,\mu$ space and the corresponding
trajectory in the thermodynamic space $S,U,N$.
If the variations in $T$ and $\mu$ are very slow,
then the trajectory in the thermodynamic space 
approaches and remains on a certain
surface which has the property of convexity and is identified as the
thermodynamic equilibrium surface. The portions of the trajectory
outside and on the surface are represented by dashed and solid
lines, respectively.}
\label{traj}
\end{figure}
%----------------------------

Let us now find the solution of the master equation.
To this end, we write the solution as
\begin{equation}
P_i(t) = P_i^*(t) + A_i(t),
\label{61}
\end{equation}
where
\begin{equation}
P_i^*(t) = \frac1{\Xi(\beta,\mu)}
\exp\{ -\beta(E_i-\mu n_i)  \},
\end{equation}
and $\Xi(\beta,\mu)$ is a time dependent quantity
such that $P_i^*(t)$ is normalized at any time,
and $A_i$ is small when compared to $P_i^*$. 
It is important to bear in mind that although
$P_i^*(t)$ obeys the relation
\begin{equation}
W_{ij}(t) P_j^*(t)=W_{ji}(t) P_i^*(t),
\end{equation}
and can be interpreted as a probability 
distribution, it is not the solution of the master equation,
given by (\ref{21}).
The substitution of $P_i^*$ on the master equation (\ref{21}) makes the
right hand side equal to zero but not the left hand side.
Replacing equation (\ref{61}) into the master equation, we get,
up to first order in the perturbation $A_i$,
\begin{equation}
\frac{d}{dt} P_i^*(t)
= \sum_j \left\{
W_{ij}(t) A_j(t) - W_{ji}(t) A_i(t)
\right\},
\label{65}
\end{equation} 

Now
\begin{equation}
\frac{d P_i^*}{d t} =
\frac{\partial P_i^*}{\partial \beta}\alpha
%\frac{\partial\beta}{d t}
+\frac{\partial P_i^*}{\partial \mu}\gamma,
%\frac{d \mu_i}{d t}
\end{equation} 
which, in view of equation (\ref{65}), 
implies that the perturbation $A_j(t)$
is of the order of $\alpha$ and $\gamma$.
From the expression (\ref{161}) for the entropy flux 
$\Phi$, it follows that $\Phi$ is also of the order 
$\alpha$ and $\gamma$.
On the other hand, if we consider the expression (\ref{9})
for the entropy production $\Pi$, it follows that $\Pi$ is of
of second order in $\alpha$ and $\gamma$.
Therefore, in the quasi-equilibrium
regime, in which we consider only terms up to first order in 
$\alpha$ and $\gamma$,
the relation $dS/dt=\Pi-\Phi$
becomes $dS/dt=-\Phi$, that is, the production of entropy
vanishes when compared with the flux of entropy.
Using this result it follows from (\ref{171}) that
the following thermodynamic relation holds
\begin{equation}
 \frac{dU}{dt} =T\frac{dS}{dt} + \mu \frac{dN}{dt}.
\label{33}
\end{equation}

Let us take a look at the thermodynamic space spanned by 
the variables $S$, $U$ and $N$. From the solution of
the master equation, we may determine these quantities 
as a function of time by using the definitions 
(\ref{6}), (\ref{11}) and (\ref{19}). The evolution
of the system may be represented by a trajectory of a point in this
space, as shown in figure \ref{traj}.
The representative
point will describe a generic trajectory in this space.
But, if $T$ and $\mu$ start to vary very slowly, the
trajectory, according to the result (\ref{33}), will
approach and remain on a certain surface of this space,
as seen in figure \ref{traj}.
According to (\ref{33}), the surface is represented by the equation
\begin{equation}
dU = TdS + \mu dN,
\label{67}
\end{equation}
so that the temperature $T$ of the thermal reservoir becomes
identified as the tangent to the surface $U(S,N)$
along the $S$ direction, $T=\partial U/\partial S$,
and can thus be interpreted as the temperature of the system.
Similarly, the chemical potential of the particle reservoir becomes
identified as the tangent to the surface $U(S,N)$
along the $N$ direction, $\mu=\partial U/\partial N$,
and can thus be interpreted as the chemical potential of the system.
Notice that, according to the inequality (\ref{34}),
this surface has the property of convexity.

We should remark that far from equilibrium, 
the temperature of the system cannot be defined
because $S$, $U$ and $N$ are not connected
by relation (\ref{67}). The
same can be said about the free energy
of systems far from equilibrium. 
Notice however that the quantity $F=U-TS$, defined
previously and called free energy, 
is not properly a property of the system because
$T$ is the temperature of the reservoir and
not the temperature of the system,
since it cannot be defined. In equilibrium
or quasi-equilibrium, however, it becomes a well defined quantity
as much as the temperature.
It is worth mentioning in addition that according to the
approach just presented, the control parameters should be
the thermodynamic variables known as thermodynamic
field variables \cite{oliveira2013}.

%--------------------------------------
\subsection{Fluxes and forces}

We consider here the contact of a system with two distinct
reservoirs. To treat this situation properly,
we assume that each pair of states $(i,j)$ is either associated
to the first reservoir or to the second reservoir or
to neither of them. In other words, the set of pairs $(i,j)$
is partitioned into three subsets, associated to the first reservoir,
to the second reservoir and neither of them, which we 
denoted by $A$, $B$ and $C$, respectively.
The transition rates associated to the reservoirs 1 and 2
are denoted by $W_{ij}^1$ and $W_{ij}^2$, respectively,
and are assumed to be
of the same form of (\ref{23}), that is, 
\begin{equation}
W_{ij}^r = K_{ij}^r 
e^{-\beta_r (E_i-E_j)/2+\beta_r\mu_r(n_i-n_j)/2},
\label{23r}
\end{equation}
for $r=1,2$, where $K_{ij}^r$ is symmetric as before.
In addition, $K_{ij}^1$ depends on $T_1$ and $\mu_1$
and is nonzero only if $(i,j)\in A$, and $K_{ij}^2$ depends on
$T_2$ and $\mu_2$ and is nonzero only if $(i,j)\in B$.
We are denoting by $T_1$ and $\mu_1$ and $T_2$ and $\mu_2$
the temperatures and chemical potentials of the two reservoirs,
and $\beta_r=1/k_BT_r$.
The full transition rate is given by 
\begin{equation}
W_{ij} = W_{ij}^0 + W_{ij}^1 + W_{ij}^2,
\end{equation}
where $W_{ij}^0$ may be nonzero only if $(i,j)\in C$.
In this case it is nonzero if
$E_i=E_j$ and $N_i=N_j$, in which case $W_{ij}^0=W_{ji}^0$.

In the following, we consider the stationary regime 
for which the stationary probability distribution $P_i$
fulfils the global balance (\ref{27}),
but not the detailed balance. In the present case
$\Pi=\Phi$ and using the expression (\ref{9f}) we may write
\begin{equation}
\Pi = k_B \sum_{r=1,2}\sum_{ij} W_{ij}^rP_j 
\ln\frac{W_{ij}^r}{W_{ji}^r}.
\label{40}
\end{equation}
where the first summation runs only over $r=1,2$.
The terms corresponding to $r=0$ vanish because
$W_{ij}^0=0$ or because $W_{ij}^0=W_{ji}^0$.
Replacing expression (\ref{23r}) 
in this equation, the entropy production can be written as
\begin{equation}
\Pi = k_B \sum_{r=1,2}\sum_{ij} W_{ij}^rP_j 
\beta_r[(E_j-E_i)-\mu_r(n_j-n_i)].
\label{41}
\end{equation}

Now, the flux of energy ${\cal J}_u$ and the flux of
particles ${\cal J}_n$ from reservoir 1 into the system
are given by
\begin{equation}
{\cal J}_u = \sum_{ij}W_{ij}^1P_j(E_i-E_j),
\label{42a}
\end{equation}
\begin{equation}
{\cal J}_n = \sum_{ij}W_{ij}^1P_j(n_i-n_j).
\label{42b}
\end{equation}
The substitution of (\ref{42a}) and (\ref{42b}) into (\ref{41})
and the use of the global balance condition (\ref{27}),
allow us to write the entropy production rate
in the bilinear form 
\cite{donder1927,onsager1931,prigogine1955}
\begin{equation}
\Pi = X_u {\cal J}_u + X_n {\cal J}_n,
\end{equation}
where $X_u$ and $X_n$ are the thermodynamic forces
\begin{equation}
X_u = \frac{1}{T_2} - \frac{1}{T_1},
\qquad\qquad
X_n = \frac{\mu_1}{T_1} - \frac{\mu_2}{T_2},
\end{equation}
conjugated to the flux of energy and particles, respectively.

%--------------------------------------
\subsection{Onsager coefficients}

When $T_2=T_1$ and $\mu_2=\mu_1$, that is, when 
$X_u=0$ and $X_n=0$, the fluxes ${\cal J}_u$ and ${\cal J}_n$
vanish. Therefore up to linear terms in $X_u$ and $X_n$
we expect the linear behavior of the fluxes,
\begin{equation}
{\cal J}_u = L_{uu}X_u + L_{un}X_n,
\end{equation}
\begin{equation}
{\cal J}_n = L_{nu}X_u + L_{nn}X_n.
\end{equation}
The coefficients $L_{uu}$, $L_{un}$, $L_{nu}$ and $L_{nn}$
are the Onsager coefficients. According to Onsager,
the cross coefficients are equal, $L_{un}=L_{nu}$,
which is the Onsager reciprocal relation.
In the following we will derive expressions for these coefficients
an prove the reciprocal relation.

We will suppose that $T_1$ and $\mu_1$ are fixed and let $T_2\to T_1$
and $\mu_2\to\mu_1$.
Let $P_i^e$ be the probability distribution corresponding
to the equilibrium case, given by
\begin{equation}
P_i^e = \frac{1}{\Xi} e^{-\beta_1 (E_i -\mu_1 n_i)}.
\end{equation}
The transition rate $W_{ij}^e$ obeys the detailed balance
\begin{equation}
W_{ij}^e P_j^e = W_{ji}^e P_i^e,
\label{43}
\end{equation}
and is given by
\begin{equation}
W_{ij}^e = K_{ij}^e e^{-\beta_1(E_i-E_j)/2+\beta_1\mu_1(n_i-n_j)/2},
\end{equation}
where $K_{ij}^e=K_{ij}^1+K_{ij}^*+K_{ij}^0$ and $K_{ij}^*$
equals $K_{ij}^2$ when $T_2\to T_1$ and $\mu_2\to\mu_1$
and $K_{ij}^0=W_{ij}^0$. 

The stationary solution $P_i$ of the master equation (\ref{27})
is written as
\begin{equation}
P_i = P_i^e(1 + a_i X_u + b_i X_n),
\label{42}
\end{equation}
up to linear term in $X_u$ and $X_n$.
Replacing into the expressions (\ref{42a}) and (\ref{42b})
we get the Onsager coefficients in the form
\begin{equation}
L_{uu} = \frac12\sum_{ij} W_{ij}^1P_j^e (a_j-a_i)(E_i-E_j),
\label{43a}
\end{equation}
\begin{equation}
L_{un} = \frac12\sum_{ij} W_{ij}^1P_j^e (b_j-b_i) (E_i-E_j),
\label{43b}
\end{equation}
\begin{equation}
L_{nu} = \frac12\sum_{ij} W_{ij}^1P_j^e (a_j-a_i) (n_i-n_j),
\label{43c}
\end{equation}
\begin{equation}
L_{nn} = \frac12\sum_{ij} W_{ij}^1P_j^e (b_j-b_i)(n_i-n_j),
\label{43d}
\end{equation}
where we have used the detailed balance condition (\ref{43}).
In the form given by equations (\ref{43b}) and (\ref{43c})
we cannot tell whether the coefficients $L_{nu}$ and $L_{un}$
are equal. Next we perform a transformation to find expressions
that will show that these coefficients are indeed equal to each other.

Replacing (\ref{42}) into (\ref{27}), and expanding the 
result up to linear terms in $X_u$ and $X_n$, we end up with
the following equations for $a_i$ and $b_i$,
\begin{equation} 
\sum_j W_{ij}^e P_j^e (a_j - a_i)
+ \frac1{k_B}\sum_j W_{ij}^*P_j^e (E_j-E_i) = 0,
\label{50a}
\end{equation}
\begin{equation} 
\sum_j W_{ij}^e P_j^e (b_j - b_i)
+ \frac1{k_B}\sum_j W_{ij}^*P_j^e (n_j-n_i) = 0,
\label{50b}
\end{equation}
where $W_{ij}^*$ equals $W_{ij}^2$ when $T_2\to T_1$ and
$\mu_2\to\mu_1$, and is given by
\begin{equation}
W_{ij}^* = K_{ij}^* e^{-\beta_1(E_i-E_j)/2+\beta_1\mu_1(n_i-n_j)/2}.
\end{equation}

Multiplying (\ref{50a}) by $E_i$ and by $a_i$ and
summing in $i$ we are lead to two equations from which
we may obtain the following expression for $L_{uu}$,
\begin{equation} 
L_{uu} = \frac1{2k_B}\sum_{ij} W_{ij}^*P_j^e (E_j-E_i)^2
-\frac{k_B}2 \sum_{ij} W_{ij}^e P_j^e (a_j-a_i)^2.
\label{52a}
\end{equation}
Multiplying (\ref{50b}) by $n_i$ and by $b_i$ and
summing in $i$ we are lead to two equations from which
we may obtain the following expression for $L_{nn}$,
\begin{equation} 
L_{nn} = \frac1{2k_B}\sum_{ij} W_{ij}^*P_j^e (n_j-n_i)^2
-\frac{k_B}2 \sum_{ij} W_{ij}^e P_j^e (b_j-b_i)^2.
\label{52b}
\end{equation}
Multiplying (\ref{50a}) by $n_i$ and (\ref{50b}) by $a_i$
and summing in $i$, we get two equations from which
we reach an expression for $L_{nu}$. Similarly, multiplying
(\ref{50a}) by $b_i$ and (\ref{50b}) by $E_i$ and summing
in $i$, we get an expression for $L_{un}$ which is equal 
to $L_{nu}$, proving the reciprocal relation. The expression
for these two quantities is given by
\[
L_{un}=L_{nu}
=\frac1{2k_B}\sum_{ij} W_{ij}^*P_j^e (E_j-E_i)(n_j-n_i)
\]
\begin{equation} 
-\frac{k_B}2 \sum_{ij} W_{ij}^e P_j^e (b_j-b_i)(a_j-a_i).
\label{52c}
\end{equation}

It is worth mentioning that in the course of derivation
of these expressions,
we have made use of the detailed balance condition, which
is thus a necessary condition to prove the reciprocal relation.
However, the expressions for the Onsager coefficients do
not depend on the equilibrium distribution alone but depend also
on the deviations $a_i$ and $b_i$.

%----------->   INICIO

\subsection{Several species of particles}

We will now treat the case of a system composed by several types of
particles in contact with two particle reservoirs, denoted by $1$ and $2$.  
In the steady state, fluxes of particles of the various types
will be established between the two reservoirs. 
Each reservoir is in fact a set of reservoirs, one for each type
of particles. As before, denoting by $E_i$ the energy of state $i$
and by $n_i^k$ the number of particles of species $k$ in state $i$,  
the rate of the transition $j\to i$ associated
to the reservoir $r$ and species $k$ is given by
\begin{equation}
W_{ij}^{rk} = K_{ij}^{rk} e^{-\beta[(E_i-E_j)-\mu_k^r(n_i^k-n_j^k)]/2},
\label{101}
\end{equation}
where $\mu_k^r$ is the chemical potential of species $k$ associated
to reservoir $r$, and $K_{ij}^{rk}$ is symmetric.  
The reservoirs are also thermal reservoirs with
a common temperature $T$ and $\beta=1/k_BT$.
A transition rate that is not associated to any
reservoir is denoted by $W_{ij}^0$ and
is assumed to be of the form
\begin{equation}
W_{ij}^0 = K_{ij}^0 e^{-\beta(E_i-E_j)/2},
\label{102}
\end{equation}
which describes the contact with a heat reservoir at temperature $T$,
where $K_{ij}^0$ is symmetric.

At the stationary state, the entropy production rate equals the
flux of entropy and is given by
\begin{equation}
\Pi = k_B \sum_{r=0,1,2}\sum_{k}\sum_{ij}W_{ij}^{rk}P_j\ln\frac{W_{ij}^{rk}}{W_{ji}^{rk}},
\end{equation}
which follows from the general expression (\ref{9f}).
The substitution of (\ref{101}) and (\ref{102})
into this expression gives
\begin{equation}
\Pi = \frac1T \sum_{r=1,2}\sum_{k}\sum_{ij}W_{ij}^{rk}P_j
\mu_k^r(n_i^k-n_j^k),
\end{equation}
where the terms involving the energy vanish.
Taking into account that the flux ${\cal J}_k$ of particles of
type $k$, from the reservoir 1 to the system, is given by
\begin{equation}
{\cal J}_k = \sum_{ij}W_{ij}^{1k}P_j(n_i^k-n_j^k),
\end{equation}
and using the total balance equation (\ref{27}), we may write
again the entropy production rate in the bilinear form
\begin{equation}
\Pi = \sum_{k} X_k {\cal J}_k,
\label{105}
\end{equation}
where 
\begin{equation}
X_k = \frac1T(\mu_k^1 -\mu_k^2)
\end{equation}
is the thermodynamic force associated to species $k$.

For the case of two types of particles, it follows from
expression (\ref{105}) that $X_1{\cal J}_1 + X_2{\cal J}_2 \geq0$
because $\Pi\geq0$.
If $X_1<0$, it is possible to have ${\cal J}_1>0$,
as long as $X_2{\cal J}_2>|X_1|{\cal J}_1$, so that
the flux of particles of type 1 will occur against the 
chemical potential gradient. This is, for instance, a mechanism for
the active transport across a cell membrane.
A simple model \cite{hill1989} of this type of transport is examined next.

\begin{figure}
\begin{center}
\epsfig{file=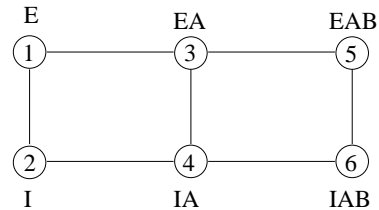,width=5cm}
\caption{\label{hillei}
Transition diagram for a model for active transport across a cell membrane.
The circles represent the possible states of a channel and the bonds represent the
possible transitions.
The possible states of the channel are:
a) open to exterior and  empty (E), or holding a molecule (EA),
or holding two molecules (EAB);
b) open to interior and empty (I), or holding a molecule (IA),
or holding two molecules (IAB).}
\end{center}
\end{figure}

A cell membrane is assumed to have a certain number of channels through which
two types of molecules may cross the membrane from the exterior
to the interior of the cell.
%The exterior and the interior of the
%cell are understood as distinct particle reservoirs.
The channels function independent of each other
so that it suffices to consider just one of them.
A channel may be open to the exterior, understood as reservoir 1,
or to the interior, understood as reservoir 2, 
and can be either empty or hold a molecule A or two molecules, one A and another B.
%The possible states of the channel are:
%(1) open to E and empty,
%(2) open to I and empty,
%(3) open to E and holding a molecule A,
%(4) open to I and holding a molecule A,
%(5) open to E and holding molecules A and B,
%(6) open to I and holding molecules A and B.
The possible states and transitions are shown in figure \ref{hillei}.

Denoting by $w_{ij}$ the rate of the transition $j\to i$,
then $w_{31}$, $w_{13}$, $w_{53}$ and $w_{35}$ are
associated to the reservoir 1, whereas  
$w_{42}$, $w_{24}$, $w_{64}$ and $w_{46}$,
associated to the reservoir 2. In this simple model,
the energies of the state are assumed to be the same
so that, according to (\ref{101}) they hold the following relations
\begin{equation}
\frac{w_{31}}{w_{13}}= e^{\beta\mu_A^1},
\qquad\qquad
\frac{w_{53}}{w_{35}}= e^{\beta\mu_B^1},
\end{equation}
\begin{equation}
\frac{w_{42}}{w_{24}}= e^{\beta\mu_A^2},
\qquad\qquad
\frac{w_{64}}{w_{46}}= e^{\beta\mu_B^2},
\end{equation}
where $\mu_A^k$ and $\mu_B^k$ are the chemical potentials of molecules
A and B associated to reservoir $k$.
The other rates are not related to the reservoirs and are symmetric,
$w_{21}=w_{12}$, $w_{43}=w_{34}$, and $w_{65}=w_{56}$. Assuming that
chemical potentials are given, the model has seven
independent transition rates.

At the stationary state, the probability distribution $P_i$
obeys the global balance equation
\begin{equation}
\sum_{j}(w_{ij}P_j - w_{ji}P_i) = 0,
\end{equation}
but do not obey the detailed balance condition, which means that
$w_{ij}P_j - w_{ji}P_i\neq 0$ in general.
The fluxes ${\cal J}_A$ and ${\cal J}_B$ of molecules A and B,
respectively, from the exterior to the interior, are given by
\begin{equation}
{\cal J}_A = w_{13}P_3-w_{31}P_1,
\end{equation}
\begin{equation}
{\cal J}_B = w_{35}P_5-w_{53}P_3,
\end{equation}
and are nonzero because detailed balance does not hold.
The entropy production rate is $\Pi=X_A{\cal J}_A+X_B{\cal J}_B$
where $X_A=(\mu_A^1-\mu_A^2)/T$ and $X_B=(\mu_B^1-\mu_B^2)/T$.
By an appropriate choice of the transition rates, it is thus possible 
to have a flux of particles $B$ against its chemical
potential gradient \cite{hill1989},
that is, it is possible to have ${\cal J}_B>0$
and $X_B<0$, as long as the flux of particles $A$ agrees with
the gradient of its chemical potential, that is, $X_A{\cal J}_A>0$.

%-------->     FIM

%----------------------------------------------------------
\section{Chemical reactions} 

%--------------------------------------
\subsection{Equilibrium}

We will be concerned in this section with a system composed by $q$
species of particles that react among themselves according
to $r$ reactions. The system is in contact with a heat reservoir
and may be closed to particles or may be open and
exchange particles with the environment. This last situation is
carried out by placing the system with particle reservoirs.
We will treat in the following the more general open case.
The results for the closed case
will readily be obtained from the results of the open case by
formally imposing the vanishing of the particle flux.

The system is placed in contact with $q$ particle reservoirs, 
one for each type of particles. Each particle reservoir is also
a thermal reservoir. The $k$-th reservoir exchanges heat,
at temperature $T$, and only particles of type $k$, at a chemical
potential $\mu_k$. Notice that all reservoirs are at the same 
temperature $T$. The number of particle 
of species $k$ in state $i$ is denoted by $n_i^k$.
If the $k$-th reservoir causes a change from 
state $j$ to state $i$ then
\begin{equation}
n_i^k\neq n_j^k  \qquad{\rm and}\qquad
n_i^{k'}=n_j^{k'},  \qquad  k'\neq k,
\label{37}
\end{equation}
because we are assuming that the $k$-th reservoir
causes a change in the number of particles of type $k$ but
causes no changes in the number of particles of the other types.

When the system is in thermodynamic equilibrium 
with the reservoirs, the probability distribution describing
the system is the Gibbs distribution 
\begin{equation}
P_i^e = \frac1\Xi e^{-\beta E_i+\beta\sum_k\mu_kn_i^k},
\label{20}
\end{equation}
where $\beta=1/k_BT$.

To set up the transition rate $\widehat{W}_{ij}^k$
describing the contact of the system with the $k$-reservoir
we use the detailed balance condition
with respect to the probability distribution (\ref{20}),
%should obey the detailed balance condition
\begin{equation}
\frac{\widehat{W}_{ij}^k}{\widehat{W}_{ji}^k}=\frac{P_i^e}{P_j^e},
\label{122}
\end{equation}
where $i$ and $j$ are states such that condition (\ref{37})
is fulfilled, so that
\begin{equation}
\frac{\widehat{W}_{ij}^k}{\widehat{W}_{ji}^k} 
= e^{-\beta(E_i-E_j) + \beta \mu_k(n_i^k-n_j^k)},
\label{31}
\end{equation}
which leads us to the following form
\begin{equation}
\widehat{W}_{ij}^k 
= \widehat{K}_{ij}^k e^{-\beta(E_i-E_j)/2 + \beta \mu_k(n_i^k-n_j^k)/2},
\label{32}
\end{equation}
where $\widehat{K}_{ij}^k$ is symmetric, that is,
$\widehat{K}_{ij}^k=\widehat{K}_{ji}^k$, and 
is positive if condition (\ref{37}) is fulfilled and vanishes otherwise.
The total transition rate $\widehat{W}_{ij}$ 
due to the contact with all reservoirs is written as the sum
\begin{equation}
\widehat{W}_{ij} = \sum_{k=1}^q \widehat{W}_{ij}^k.
\label{10}
\end{equation}
Notice that at most one of the $q$ terms on the right hand side can
be nonzero.

Let us consider now the occurrence of chemical reactions.
The number of particles of each 
species will vary not only because of the contact with
the reservoirs but also because of the reactions.
We consider the occurrence of $r$ reactions
described by the chemical equations
\begin{equation}
\sum_{k=1}^q\nu_{k\ell}B_k = 0,   \qquad\qquad \ell=1,2,\ldots,r,
\label{140}
\end{equation}
where $B_k$ denotes the chemical formula of species
$k$ and $\nu_{k\ell}$ are the stoichiometric coefficients,
which are negative for the reactants and positive for the
products of the reaction. 
If the $\ell$-th reaction causes a change from state $j$ to 
state $i$ then
\begin{equation}
n_i^k-n_j^k = \nu_{k\ell} \qquad{\rm or}\qquad
n_i^k-n_j^k = - \nu_{k\ell}.
\label{14}
\end{equation}

To set up the transition rate $\widetilde{W}_{ij}^\ell$ 
describing the change caused by the $\ell$-th reaction 
we assumed that it obeys the Arrhenius equation
\cite{arrhenius1889,moore1965}
\begin{equation}
\frac{\widetilde{W}_{ij}^\ell}{\widetilde{W}_{ji}^\ell} = e^{-\beta(E_i-E_j)}.
\label{121}
\end{equation}
The most general form of the transition rate is
\begin{equation}
\widetilde{W}_{ij}^\ell 
= \widetilde{K}_{ij}^\ell e^{-\beta(E_i-E_j)/2},
\end{equation}
where the prefactor is symmetric, that is, 
$\widetilde{K}_{ij}^\ell=\widetilde{K}_{ji}^\ell$,
and is positive or vanishes according to whether the condition 
(\ref{14}) is fulfilled or not. The transition rate 
$\widetilde{W}_{ij}$ due to all reactions is written as the sum
\begin{equation}
\widetilde{W}_{ij} = \sum_{\ell=1}^r \widetilde{W}_{ij}^\ell.
\label{10b}
\end{equation}
Notice that at most one of the $r$ terms on the right hand side can
be nonzero.

The full transition rate $W_{ij}$ describing the $r$ reactions as
well as the contact with the $q$ reservoirs is given by
\begin{equation}
W_{ij} = \widetilde{W}_{ij} + \widehat{W}_{ij}.
\label{10a}
\end{equation}
Again, just one of the two terms on the right hand side can
be nonzero.

In equilibrium, detailed balance should be obeyed for each
one of the transition rates on the right-hand side of (\ref{10a}).
We have seen that this is the case of the transition
rates $\widehat{W}_{ij}^k$, related to the contact with each
reservoir, when the probability distribution is that given by (\ref{20}).
It suffices therefore to impose detailed balance to the
transition rate associated to each chemical reaction.
To this end we compare the ratio (\ref{121}) with the ratio
\begin{equation}
\frac{P_i^e}{P_j^e} 
= e^{-\beta(E_i-E_j) + \beta\sum_k \nu_{k\ell}\mu_k},
\end{equation}
obtained from (\ref{20}) and
valid when the first of the two conditions in (\ref{14})
is fulfilled.
The condition of detailed balance is obeyed
when the two ratios are equal to each other, that is, when
\begin{equation}
\sum_k \nu_{k\ell}\mu_k = 0,
\label{38}
\end{equation}
for each reaction $\ell$. The same conclusion is obtained if we use
the second of the two conditions in (\ref{14}). Equation (\ref{38})
is the well known equilibrium condition that should
be fulfilled when chemical reactions take place in a system
\cite{callen1960,oliveira2013}.
In the presence of chemical reactions and in equilibrium, 
the chemical potentials of the chemical species cannot be 
independent but are related by (\ref{38}).
In other words, the equilibrium occurs only when the 
chemical potentials $\mu_k$ of the particle reservoirs are
tuned so that (\ref{38}) is fullfiled. Otherwise,
the system will be out of equilibrium, as we shall see next.

%--------------------------------------
\subsection{Nonequilibrium regime}

Let us now suppose that the condition (\ref{38}) is not
obeyed. In this case the detailed balance condition
is not fulfilled and the system cannot be in 
equilibrium. Each reaction is shifted either to the products
or to the reactants. That is, for a given reaction,
either the products are  
being created and the reactants being annihilated (forward reaction)
or the reactants are being created and the products being annihilated 
(backward reaction). In this nonequilibrium regime 
the time variation in the number of particles
has two parts: one due to the flux of particles from the
reservoirs and the other due the creation and annihilation
caused by the reactions. 

Using the master equation (\ref{21}), 
we see that the average number of particles $N_k$ of type $k$, 
\begin{equation}
N_k(t) = \sum_i n_i^k P_i(t),
\end{equation}
evolves as
\begin{equation}
\frac{dN_k}{dt} = \sum_{ij}W_{ij}P_j(n_i^k-n_j^k).
\label{71}
\end{equation}
According to (\ref{10a}), the transition rate $W_{ij}$ has two parts,
one related to the reservoirs, which is $\widehat{W}_{ij}$, and the
other related to the chemical reactions, which is $\widetilde{W}_{ij}$,
so that (\ref{71}) can be written in the form
\begin{equation}
\frac{dN_k}{dt} = \sum_{ij}\widetilde{W}_{ij}P_j(n_i^k-n_j^k) + \Phi_k,
\label{157}
\end{equation}
where $\Phi_k$ is given by
\begin{equation}
\Phi_k = \sum_{ij}\widehat{W}_{ij} P_j(n_i^k-n_j^k),
\label{39}
\end{equation}
and describes the flux of
particle from the $k$-th reservoir to the system. 
The contact of the system with the $k$-th reservoir,
described by the transformation (\ref{37}),
causes no changes in $n_i^{k'}$, $k'\neq k$. As a consequence,
\begin{equation}
\widehat{W}_{ij}^k(n_i^{k'}-n_j^{k'}) = 0,
\qquad\qquad k\neq k',
\label{60}
\end{equation}
Using (\ref{10}) and the result (\ref{60}), the flux of
particle (\ref{39}) is written as
\begin{equation}
\Phi_k = \sum_{ij}\widehat{W}_{ij}^k P_j(n_i^k-n_j^k).
\label{39a}
\end{equation}

The summation in the right-hand side of (\ref{157})
describes the change in the number of particle due to
the chemical reactions.
To describe properly this part, 
which corresponds to the creation and annihilation
of particles caused by the reactions,
it is convenient to use a new set of variables
in the place of the set of variables $n_i^k$,
$k=1,2,\ldots,q$. 
The new variables are denoted by $\sigma_i^\ell$, $\ell=1,2,\ldots,r$
and $x_i^\ell$, $\ell=r+1,r+2,\ldots,q$ and defined by
the linear transformation
\begin{equation}
n_i^k = x_i^k + \sum_{\ell=1}^r \nu_{k\ell}\sigma_i^\ell,
\qquad k=1,2,\ldots,q,
\label{29}
\end{equation}
where the quantities $x_i^k$, $k=1,2\ldots,r$ are not variables
but arbitrary constants chosen to be the same for all $i$.
If a transformation $n_i^k\to n_j^k$
is performed according to the $\ell$-th chemical reaction
(\ref{140}), described by the transformation (\ref{14}),
the variables $x_i^k$ remains unchanged,
that is, $x_j^k=x_i^k$. As a consequence of this invariance,
\begin{equation}
\widetilde{W}_{ij}^\ell(x_i^k-x_j^k) = 0.
\label{58}
\end{equation}
In addition, according to the transformation (\ref{14}),
the variables $\sigma_i^m$, $m\neq\ell$,
associated to the other reactions remain 
unchanged, $\sigma_j^m=\sigma_i^m$, and, as a consequence,
\begin{equation}
\widetilde{W}_{ij}^m(\sigma_i^\ell-\sigma_j^\ell) = 0,
\qquad\qquad m\neq\ell.
\label{59}
\end{equation}

Using (\ref{29}) and the results (\ref{58}) and (\ref{59}),
we obtain 
\begin{equation}
\sum_{ij}\widetilde{W}_{ij} P_j(n_i^k-n_j^k) = 
\sum_\ell\nu_{k\ell} \chi_\ell,
\label{135}
\end{equation}
where the quantity $\chi_\ell$ is
\begin{equation}
\chi_\ell = 
\sum_{ij}\widetilde{W}_{ij}^\ell P_j(\sigma_i^\ell-\sigma_j^\ell).
\label{156}
\end{equation}
The variation in the number of particles can then be written as
\begin{equation}
\frac{dN_k}{dt} = \sum_\ell \nu_{k\ell}\chi_\ell +  \Phi_k.
\label{46}
\end{equation}

When $\chi_\ell>0$, the $\ell$-th chemical reaction is shifted 
to the right, in the direction of the products.
When $\chi_\ell<0$, it is shifted to the leftt, 
in the direction of the reactants.
The extent of reaction $\xi_\ell$ is defined as the average
of $\sigma_i^\ell$,
\begin{equation}
\xi_\ell = \sum_i \sigma_i^\ell P_i.
\end{equation}
From the master equation and using the properties (\ref{60})
and (\ref{59}), we get
\begin{equation}
\frac{d\xi_\ell}{dt} = \chi_\ell,
\end{equation}
and we may conclude that the quantity $\chi_\ell$ is the rate of the
extent of the $\ell$-th reaction.

The time variation of the internal energy is written as
\begin{equation}
\frac{dU}{dt} = \sum_{ij} W_{ij}P_j(E_i-E_j). 
%\Phi_u,
\label{141}
\end{equation}
%where $\Phi_u$ is given by
%\begin{equation}
%\Phi_u = \sum_{ij} W_{ij}P_j(E_i-E_j). 
%\label{5a}
%\end{equation}

Let us now consider the time variation of entropy
\begin{equation}
\frac{dS}{dt} = \Pi - \Phi,
\label{45}
\end{equation}
where the flux $\Phi$ is given by (\ref{9}) which,
by the use of (\ref{10a}), (\ref{10b}) and (\ref{10}),
is given by
\[
\Phi = k_B \sum_{ij}\sum_k\widehat{W}_{ij}^kP_j
\ln\frac{\widehat{W}_{ij}^k}{\widehat{W}_{ji}^k}
\]
\begin{equation}
+ k_B\sum_{ij}\sum_\ell\widetilde{W}_{ij}^\ell P_j
\ln\frac{\widetilde{W}_{ij}^\ell}{\widetilde{W}_{ji}^\ell}.
\end{equation}

Substituting the rates (\ref{31}) and (\ref{121}) 
into this equation we get
\[
\Phi = k_B\sum_{ij} W_{ij}P_j(-\beta)(E_i-E_j)
\]
\begin{equation}
+ k_B\sum_{ij}\sum_k \widehat{W}_{ij}^k P_j\beta\mu_k(n_i^k-n_j^k).
\end{equation}
Using equations (\ref{141}) and
%the definition of $\Phi_u$, given by (\ref{5a}), and the equation
(\ref{39a}) we may write the flux of entropy as
\begin{equation}
%\Phi = - \frac{1}{T}\Phi_u + \frac{1}{T}\sum_k \mu_k\Phi_k.
\Phi = - \frac{1}{T}\frac{dU}{dt} + \frac{1}{T}\sum_k \mu_k\Phi_k.
\end{equation}
Substituting into (\ref{45}) and taking into account the
equations (\ref{141}), we get
\begin{equation}
\frac{dS}{dt} = \Pi + \frac{1}{T}\frac{dU}{dt} 
- \frac{1}{T}\sum_k \mu_k\Phi_k.
% --> ATENCAO --> o sinal aqui e' negativo! <---------------------
\label{66}
\end{equation}
Using (\ref{46}), we reach the result
\begin{equation}
\frac{dS}{dt} = \Pi + \frac{1}{T}\frac{dU}{dt}
- \sum_\ell\frac{A_\ell}{T}\chi_\ell 
- \frac{1}{T}\sum_k \mu_k\frac{dN_k}{dt},
\end{equation}
where $A_\ell$ is the De Donder affinity \cite{donder1927}
\begin{equation}
A_\ell = - \sum_k \nu_{k\ell}\mu_k,
\label{111}
\end{equation}
associated to the $\ell$-th chemical reaction.

In the stationary state $dS/dt=0$, $dU/dt=0$, and $dN_k/dt=0$,
and we reach the following expression for the production
of entropy in the stationary state 
\cite{donder1927,prigogine1955,glansdorff1971,nicolis1977}
\begin{equation}
\Pi = \sum_\ell\frac{A_\ell}{T}\chi_\ell,
\label{113}
\end{equation}
or, in the equivalent form, 
\begin{equation}
\Pi = \sum_\ell\frac{A_\ell}{T}\frac{d\xi_\ell}{dt},
\end{equation}
equation originally introduced by De Donder \cite{donder1927}. 
In equilibrium there is no
production of entropy and the affinities vanish, 
$A_\ell=0$, in accordance with (\ref{38}), and
the rate in which the $\ell$-th reaction proceeds vanish
as well, $\chi_\ell=d\xi_\ell/dt=0$.

%material acrescentado --->
In a nonequilibrium stationary state, a flux
of particles is continuously taking place, which
sustain the chemical reactions. The quantities $\Phi_k$
and $\chi_\ell$ are nonzero, in general. At the same time there
is a flux of heat toward the system, characterizing
an endothermic reaction, or from the system, characterizing
en exothermic reaction. To understand this situation
we write down the variation in the energy, given by 
equation (\ref{141}) in the form
\begin{equation}
\frac{dU}{dt} = \sum_\ell R_\ell + \Phi_u,
\label{141a}
\end{equation}
where $R_\ell$ is the energy delivered by the $\ell$-th reaction per unit time,
given by, 
\begin{equation}
R_\ell = \sum_{ij} \widetilde{W}_{ij}^\ell P_j(E_i-E_j),
\label{141b}
\end{equation}
and $\Phi_u$ is the heat flux to the system, given by
\begin{equation}
\Phi_u = \sum_{ij} \sum_k \widehat{W}_{ij}^kP_j(E_i-E_j).
\label{141c}
\end{equation}
In the stationary state $dU/dt=0$ and $\Phi_u=-\sum_\ell R_\ell$.
If $\Phi_u<0$, the reactions are exothermic. If $\Phi_u>0$,
they are endothermic. Notice that there is no contribution to
the entropy production rate coming from the flux of heat
because the temperatures of the reservoirs are the same.

%novo material
Although we have considered a system in contact with one
reservoir for each type of particles, the formulas can
easily be adapted to the case in which the system is closed
to some types of particles. If the system is closed to particles
of type $k$, then it suffices to formally set $\mu_k=0$
and $\widehat{W}_{ij}^k=0$ so that $\Phi_k=0$ for this species.
%fim do novo material
It is worth mentioning that in the case of a closed system,
when there is no flux of particles from the environment,
$\Phi_k=0$ for all species, and using equation (\ref{66}), we see that
$dF/dt=-T\Pi$ where $F=U-TS$ is the free energy,
so that $dF/dt\leq 0$.
Therefore, the chemical reactions occur in a direction
such that the variations in the number of particles
will decrease the free energy \cite{lewis1923}.

%material novo ------->
%{Michaelis-Menten mechanism}

As an example of the approach just developed
we analyze a system with four species of 
particles and two reactions, which are 
\begin{equation}
B_1 + B_2 = B_3, \qquad\qquad B_3 = B_2 + B_4,
\end{equation}
and represent the Michaelis-Menten mechanism in which a
the substrate $B_1$ is converted, in two steps, into the product $B_4$
by the action of an enzyme. The substrate $B_1$ reacts
with the enzyme $B_2$ giving rise to a complex $B_3$
which in turn breaks up into the product $B_4$
and the enzyme $B_2$. It is assumed that both reactions
have reverses. The system is assumed to be closed to the particles $B_2$
and $B_3$, and is in contact with reservoirs
of particles of type $B_1$ and $B_4$. 

Using formula (\ref{111}), and bearing in mind that we should
set $\mu_2=0$ and $\mu_3=0$ in this formula,
the affinities $A_1 $ and $A_2$ associated to the two reactions are given by
\begin{equation}
A_1 = \mu_1, \qquad\qquad A_2 =-\mu_4.
\end{equation}
The variations in the number of particles of each species are
\begin{equation}
\frac{dN_1}{dt} =-\chi_1 + \Phi_1,
\qquad\qquad
\frac{dN_2}{dt} =-\chi_1 + \chi_2,
\end{equation}
\begin{equation}
\frac{dN_3}{dt} = \chi_1 -\chi_2,
\qquad\qquad
\frac{dN_4}{dt} = \chi_2 + \Phi_4.
\end{equation}

In the stationary state,
$\chi_1=\chi_2=\Phi_1=-\Phi_4$, so that, using (\ref{113}),
the entropy production rate is found to be
\begin{equation}
\Pi = \frac{\chi_1}{T}(\mu_1 - \mu_4). 
\end{equation}
We may now draw the following conclusion for the
case of a nonequilibrium steady state situation, for
which $\Pi>0$. If $\mu_1>\mu_4$ then $\chi_1>0$
and $\chi_4>0$ so that the two reaction equations are
shifted to the right establishing a continuous
annihilation of particles of type $B_1$,
which come from reservoir $B_1$ because $\Phi_1>0$,
and production of particles of type $B_4$,
which go to reservoir $B_4$ because $\Phi_4<0$.

%Ate aqui ---->

%--------------------------------------
\subsection{Onsager coefficients}

In the nonequilibrium stationary state but close to equilibrium we may
expand the rates of the extents of reaction  
$\chi_\ell$ in terms of the affinities $A_\ell$ to get
\begin{equation}
\chi_\ell = \sum_{m} L_{\ell m}A_m,
\end{equation}
where $L_{\ell m}$ are the Onsager coefficients. They obey the
reciprocal relations, which we demonstrate next.

We start by expanding the stationary probability distribution
$P_i$, that satisfies the global balance equation (\ref{27}),
around the equilibrium distribution $P_i^e$ given by
\begin{equation}
P_i^e = \frac1Z e^{-\beta E_i+\beta\sum_k\mu_k^*n_i^k},
\label{20a}
\end{equation}
where the chemical potentials $\mu_k^*$ obey the 
equilibrium condition
\begin{equation}
\sum_k \nu_{k\ell}\mu_k^* = 0.
\label{87}
\end{equation}
We assume an expansion of the form
\begin{equation}
P_i = P_i^e(1 + \sum_{\ell=1}^r  R_{i\ell} A_\ell +
\sum_{k=r+1}^q a_{ik} \Delta\mu_k),
\label{32a}
\end{equation}
where $\Delta\mu_k = \mu_k-\mu_k^*$.

We also expand $\widehat{W}_{ij}^k$, given by (\ref{32}),
around its value at equilibrium
\begin{equation}
\widehat{W}_{ij}^{*k} 
= \widehat{K}_{ij}^{k} e^{-\beta(E_i-E_j)/2+\beta \mu_k^*(n_i^k-n_j^k)/2},
\end{equation}
to get
\begin{equation}
\widehat{W}_{ij}^k 
= \widehat{W}_{ij}^{*k} \{1+\beta \Delta\mu_k(n_i^k-n_j^k)/2 \}.
\label{32b}
\end{equation}
The transition rate $\widetilde{W}_{ij}^\ell$ needs no expansion 
because this quantity is also its value at equilibrium
since it does not depend on the chemical potentials.

Replacing the expansions (\ref{32a}) and (\ref{32b})
into the global balance equation (\ref{27}), we get 
\[
\sum_j W_{ij}^* P_j^e \sum_{\ell=1}^r A_\ell(R_{j\ell} - R_{i\ell})
\]
\[
+\sum_j W_{ij}^* P_j^e \sum_{k=r+1}^q (a_{jk}-a_{ik}) \Delta\mu_k 
\]
\begin{equation}
+\sum_j \sum_k
\widehat{W}_{ij}^*P_j^e \beta \Delta\mu_k(n_i^k-n_j^k) = 0,
\label{88}
\end{equation}
where
\begin{equation}
\widehat{W}_{ij}^* = \sum_k \widehat{W}_{ij}^{*k},
%\end{equation}
\qquad{\rm and}\qquad
%\begin{equation}
W_{ij}^* = \widehat{W}_{ij}^* + \widetilde{W}_{ij}.
\end{equation}
Using (\ref{29}) and taking into account relation (\ref{87}),
we see that
\[
 \sum_k \Delta\mu_k(n_i^k-n_j^k) =
\]
\begin{equation}
=\sum_{k=r+1}^q \Delta\mu_k(x_i^k-x_j^k) 
- \sum_{\ell=1}^r A_\ell (\sigma_i^\ell-\sigma_j^\ell), 
\end{equation}
which is replaced in (\ref{88}) to get an expression
linear in $A_\ell$ and $\Delta\mu_k$. 
Since the coefficients of $A_\ell$ and $\Delta\mu_k$
in this expression should vanish, we obtain
\begin{equation}
\sum_j W_{ij}^* P_j^e (a_{jk}-a_{ik}) +
\sum_j \widehat{W}_{ij}^{*}P_j^e(x_i^k-x_j^k) = 0,
\end{equation}
valid for $r+1\leq k\leq q$,
\begin{equation}
\sum_j W_{ij}^* P_j^e (R_{j\ell} - R_{i\ell})
-\beta\sum_j \widehat{W}_{ij}^{*}P_j^e (\sigma_i^\ell-\sigma_j^\ell) = 0,
\label{62}
\end{equation}
valid for $1\leq\ell\leq r$.
These last two equations determine $a_{ik}$ and $R_{i\ell}$.

Let us consider now the expansion of the rate of
the extent of reaction $\chi_\ell$, given by (\ref{156}).
Replacing the expansion (\ref{32a}) into (\ref{156}),
we get an expression linear in $A_m$ and $\Delta\mu_k$.
The coefficient of $A_m$ is the Onsager coefficient $L_{\ell m}$
which is given by
\begin{equation}
L_{\ell m} = \sum_{ij}\widetilde{W}_{ij} P_j^e (R_{jm}-R_{im})
(\sigma_i^\ell-\sigma_j^\ell).
\label{44}
\end{equation}

Next we use equation (\ref{62}) to write the Onsager
coefficient in a more appropriate form. To this end
we proceed as follows. We multiply (\ref{62}) by $\sigma_i^m$
and sum in $i$ to get a first equation. Next, we
multiply (\ref{62}) by $R_i^m$ and sum in $i$ to get 
a second equation. From these two equations we get an
equation for the right-hand side of (\ref{44}) from which
we reach the following expression 
\[
L_{\ell m}
= \frac{\beta}2\sum_{ij} \widehat{W}_{ij}^*P_j^e(\sigma_i^m-\sigma_j^m)
(\sigma_i^\ell -\sigma_j^\ell)
\]
\begin{equation}
- \frac{1}{2\beta}\sum_{ij} W_{ij}^* P_j^e (R_{i\ell}
- R_{j\ell})(R_{im} -R_{jm}).
\end{equation}
From this expression it follows that
\begin{equation}
L_{m\ell} = L_{\ell m},
\end{equation}
which is the Onsager reciprocal relation \cite{onsager1931}.

%----------------------------------------------------------
\section{Fokker-Planck equation}

%--------------------------------------
\subsection{Langevin equations}

In this section we are concerned with systems that follow a continuous
time Markovian process in the continuous state space, the phase space.
We consider a system of particles that follows a
dynamics described by the following set of Langevin equations,
interpreted according to It\^o,
\begin{equation}
m\frac{dv_i}{dt} = F_i(x) -\alpha_i v_i + \mathfrak{F}_i(t),
\label{387}
\end{equation}
where $m$ is the mass of each particle, $v_i=dx_i/dt$ and
$x_i$ is the position of the $i$-th particle
and $x$ denotes the vector $x=(x_1,\ldots,x_n)$.
We will also use the notation $v=(v_1,\ldots,v_n)$.
The quantity $F_i(x)$ is the force
acting on the $i$-th particle, and $\mathfrak{F}_i(t)$
is a stochastic variable with the properties
\begin{equation}
\langle\mathfrak{F}_i(t)\rangle = 0,
\end{equation}
\begin{equation}
\langle\mathfrak{F}_i(t)\mathfrak{F}_j(t')\rangle
= 2B_{ij} \delta(t-t'),
\end{equation}
where $B_{ij}$ may depend on $x$ and $v$.

Notice that we are considering the so called underdamped systems,
for which the state of a particle is defined by its position and
velocity \cite{tome2010},
in opposition to the overdamped case, for which
the state of a particle is defined only by its position
\cite{tome2006}. 

The quantities $\mathfrak{F}_i(t)$ are random forces acting on
the particles including the ones that describe the
contact of the system with the environment. We will
treat two cases: one in which the system is isolated
(microcanonical ensemble) and the other in which the
system is in contact with a heat reservoir (canonical 
ensemble). In the first case the forces $F_i$ are conservative
and the stochastic forces
are set up in such a way that the energy is conserved
in any stochastic trajectory. In thermodynamic 
equilibrium they will lead to the Gibbs microcanonical 
probability distribution. In the second case the forces $F_i$
are also conservative and the random forces
are set up in such a way that in thermodynamic equilibrium
they will lead to the Gibbs canonical distribution.

Using the It\^o interpretation, we can show that the Langevin
equations (\ref{387}) are associated to the following Fokker-Planck equation
\[
\frac{\partial P}{\partial t} = 
-\sum_i \frac{\partial}{\partial x_i}(v_iP)
-\frac1{m}\sum_i \frac{\partial}{\partial v_i}(F_i P)
\]
\begin{equation}
+\sum_i\frac{\alpha_i}{m}\frac{\partial}{\partial v_i}(v_iP)
+\frac1{m^2}\sum_{ij}
\frac{\partial^2}{\partial v_i\partial v_j}(B_{ij}P),
\end{equation}
equation that gives the time evolution of the
probability distribution $P(x,v,t)$ of $x$ and $v$ at time $t$.
It is convenient to write down the Fokker-Planck equation in 
the following form
\begin{equation}
\frac{\partial P}{\partial t} = - \sum_i\left(K_i + 
\frac{\partial J_i}{\partial v_i} \right), 
\label{348}
\end{equation}
where $K_i$ and $J_i$ are given by
\begin{equation}
K_i = v_i \frac{\partial P}{\partial x_i}
+ \frac{F_i}{m}\frac{\partial P}{\partial v_i},
\label{346}
\end{equation}
and
\begin{equation}
J_i = -\frac{\alpha_i}{m}v_iP
-\frac1{m^2} \sum_j \frac{\partial}{\partial v_j}(B_{ij}P).
\end{equation}

Let us consider now the time variation of entropy $S$, given by
\begin{equation}
S(t) = -k_B \int P(x,v,t)\ln P(x,v,t) dxdv.
\label{352}
\end{equation}
The derivative of $S$ gives
\begin{equation}
\frac{dS}{dt} = -k_B \int \left(\frac{\partial P}{\partial t}\right)\ln P dxdv.
\label{353}
\end{equation}
After replacing (\ref{348}) into this equation
and performing appropriate integrations by parts we reach
the following expression
\begin{equation}
\frac{dS}{dt} = -k_B\sum_i\int 
\frac{J_i}{P}\left(\frac{\partial P}{\partial v_i}\right) dxdv.
\label{354}
\end{equation}
The terms corresponding to $K_i$ vanish, that is,
\begin{equation}
-k_B\sum_i\int K_i \ln P dxdv = 0.
\label{444}
\end{equation}
We are assuming that $P$ and its derivatives vanish at the 
boundary of integration. 

%--------------------------------------
\subsection{Microcanonical ensemble}

Here we treat the case of an isolated system,
with no contact with the environment so that
the energy is strictly conserved. We thus
assume that the force $F_i$ are conservative
so that $F_i=-\partial V/\partial x_i$ which allows
us to define the energy function $E(v,x)$ as
\begin{equation}
E(x,v) = \sum_i \frac{m}2 v_i^2 + V(x).
%\label{325}
\end{equation}
The strict conservation of energy means to say that
$E(x,v)$ should be
a constant along any stochastic trajectory in phase space.
This condition is fullfiled by the following
set of Langevin equations, understood
according the Stratonovich interpretation,
\begin{equation}
m\frac{dv_i}{dt} = F_i(x) + \sum_{j(\neq i)}\xi_{ij} v_j,
\label{345}
\end{equation}
where $\xi_{ij}$ are stochastic variables with the antisymmetric
property $\xi_{ji}=-\xi_{ij}$. 
% material adicionado
The multiplicative noise at the
right-hand side changes the velocities of the particles while
keeping the kinetic energy invariant and can be interpreted
as random elastic collisions of the particles with themselves 
or with immobile scatters.
A similar noise has been used do describe a particle that moves
at constant speed but changes direction at random times
\cite{lemons2002,landi2014}.
% ate aqui

Multiplying (\ref{345}) by $v_i$ and summing in $i$ we 
may conclude, after using the antisymmetric relation
$\xi_{ji}=-\xi_{ij}$, that $E(v,x)$ is strictly
conserved along any stochastic path $x(t)$, $v(t)$. 
Therefore, the equation of motion (\ref{345}) describes
a system of particles evolving in time in such a way
that the energy is strictly constant.
In analogy with equilibrium statistical mechanics,
this defines a microcanonical ensemble.

The stochastic variables $\xi_{ij}(t)$ are defined by the
relations
\begin{equation}
\langle \xi_{ij}(t)\rangle = 0,
\end{equation}
and
\begin{equation}
\langle \xi_{ij}(t)\xi_{ij}(t')\rangle = 2\lambda_{ij} \delta(t-t'),
\end{equation}
where $\lambda_{ij}\geq0$ is a parameter that gives the strength
of the stochastic noise.
Using the Stratonovich interpretation, and taking into account
the antisymmetric property $\xi_{ji}=-\xi_{ij}$ of the noise,
we may write down the
associate Fokker-Planck equation, given by
\[
\frac{\partial P}{\partial t} = 
-\sum_i \frac{\partial}{\partial x_i}(v_iP)
-\frac1{m}\sum_i \frac{\partial}{\partial v_i}(F_i P)
\]
\begin{equation}
+\frac1{m^2}\sum_{ij}  \lambda_{ij}
\left(
 v_j\frac{\partial}{\partial v_i}v_j\frac{\partial P}{\partial v_i}
-v_j\frac{\partial}{\partial v_i}v_i\frac{\partial P}{\partial v_j}
\right),
\label{349}
\end{equation}
equation that gives the time evolution of the
probability distribution $P(x,v,t)$ of $x$ and $v$ at time $t$.
The last summation extends over $i\neq j$ and we recall that
$\lambda_{ji}=\lambda_{ij}\geq0$.

It is worth mentioning that equation (\ref{345}), understood
in the Stratonovich sense, is equivalent to the following
equation, interpreted according to It\^o interpretation, 
\begin{equation}
m\frac{dv_i}{dt} = F_i(x) - \alpha_i v_i + 
\sum_{j(\neq i)}\xi_{ij} v_j,
\label{345ito}
\end{equation}
where
\begin{equation}
\alpha_i = \sum_{j(\neq i)} \lambda_{ij}.
\end{equation}
Of course, this equation leads to the same Fokker-Planck equation
(\ref{349}).

It is convenient to write down the Fokker-Planck equation in 
the form given by (\ref{348})
where $K_i$ is given by (\ref{346}) 
and $J_i$ is given by
\begin{equation}
J_i = \sum_{j(\neq i)} J_{ij}v_j,
\label{367}
\end{equation}
\begin{equation}
J_{ij} = \frac1{m^2} \lambda_{ij}\left(
 v_i\frac{\partial P}{\partial v_j}
-v_j\frac{\partial P}{\partial v_i}
\right),
\end{equation}

Let us determine now the time derivative of entropy,
which is given by equation (\ref{354}). 
After replacing (\ref{367}) into equation (\ref{354})
and performing appropriate integration by parts we reach
the following expression
\begin{equation}
\frac{dS}{dt} = \frac{k_B}{m^2} \sum_{i<j}\lambda_{ij}\int \frac{1}{P}
\left(v_j\frac{\partial P}{\partial v_i} - v_i\frac{\partial P}{\partial v_j}
\right)^2 dxdv.
\label{413}
\end{equation}
We are assuming that $P$ and its derivatives vanish at the 
boundary of integration. The right-hand side of this equation
is clearly nonnegative and is therefore identified as the
entropy production rate
\begin{equation}
\Pi = \frac{k_B}{m^2} \sum_{i<j}\lambda_{ij} \int \frac{1}{P}
\left(v_j\frac{\partial P}{\partial v_i} - v_i\frac{\partial P}{\partial v_j}
\right)^2 dxdv,
\label{350}
\end{equation}
which can also be written in the form
\begin{equation}
\Pi = k_B \sum_{i<j}\frac{m^2}{\lambda_{ij}}
\int \frac{J_{ij}^2}{P} dxdv,
\label{350a}
\end{equation}
where the summation is over $ij$ such that $\lambda_{ij}\neq0$,
so that
\begin{equation}
\frac{dS}{dt} = \Pi.
\end{equation}
In the present case there is no entropy flux, 
\begin{equation}
\Phi = 0,
\label{350b}
\end{equation}
which is consistent
with our interpretation that equations (\ref{345}) describe an
isolated system. Taking into account that $\Pi\geq0$ it follows
at once that $dS/dt\geq0$ for an isolated system.

In the stationary state, which is a thermodynamic equilibrium,
the probability distribution $P^e(x,v)$ depends on $x$ and $v$
only through $E(x,v)$, that is, $P^e(x,v)$ is a function of $E(x,v)$.
This statement can be checked by substitution on the right-hand
side of the Fokker-Planck equation (\ref{349}). Since
$E(x,y)$ is invariant along any path in phase space and 
supposing that initially its value is $U$, it follows that
\begin{equation}
P^e(x,v) = \frac1\Omega \delta (U-E(x,v)),
\end{equation}
where $\Omega$ is a normalization constant that depends on $U$.
We remark that in this case $\Pi$, given by (\ref{350}), vanishes, as
expected.

%--------------------------------------
\subsection{Canonical ensemble}

Now we consider the case of a system in contact with
a heat reservoir. 
In fact, we will consider the more general case in 
which each particle $i$ is in contact with a reservoir
at temperature $T_i$. 
The appropriate set of Langevin equations that describes
this situation is given by
\begin{equation}
m\frac{dv_i}{dt} = F_i(x) -\alpha_i v_i + \zeta_i(t),
\label{305}
\end{equation}
where $\zeta_i(t)$ is a stochastic variable with the properties
\begin{equation}
\langle\zeta_i(t)\rangle = 0,
\end{equation}
\begin{equation}
\langle\zeta_i(t)\zeta_j(t')\rangle
= 2\alpha_i k_B T_i\,\delta_{ij}\delta(t-t'),
\end{equation}
where $T_i$ and $\alpha_i$ are parameters.
The two last terms in equation (\ref{305})
are interpreted as describing the contact of the
$i$-th particle with the heat bath at a temperature $T_i$
and $\alpha_i$ is the strength of the interaction with
the heat reservoir.

To the set of Langevin equations (\ref{305}) is associated
the Fokker-Planck equation
\[
\frac{\partial P}{\partial t} = 
-\sum_i \frac{\partial}{\partial x_i}(v_iP)
-\frac1{m}\sum_i \frac{\partial}{\partial v_i}(F_i P)
\]
\begin{equation}
+\frac1m\sum_i \alpha_i \frac{\partial}{\partial v_i}(v_iP)
+\frac{k_B}{m^2}\sum_i \alpha_i T_i \frac{\partial^2 P}{\partial v_i^2},
\label{315}
\end{equation}
equation that gives the time evolution of the
probability distribution $P(x,v,t)$ of $x$ and $v$ at time $t$.

The Fokker-Planck equation can again be written in 
the form given by (\ref{348}) where $K_i$ is given by (\ref{346}) 
and $J_i$ is given by
\begin{equation}
J_i = -\frac{\alpha_i v_i}{m}P - \frac{\alpha_i k_BT_i}{m^2}
\frac{\partial P}{\partial v_i}.
\label{340}
\end{equation}

Again the derivative of entropy is given by (\ref{354}). Replacing (\ref{340})
into (\ref{354}) we get the following expression \cite{tome2010}
\begin{equation}
\frac{dS}{dt} = \sum_i \int\left( 
\frac{m^2}{\alpha_i T_i} \frac{J_i^2}{P} + \frac{m}{T_i} v_i J_i 
\right)dxdv.
\label{359}
\end{equation}
The summation in (\ref{359}) extends only
to the terms for which $\alpha_i\ne0$ and $T_i\neq0$.

The first term on the right-hand side of equation (\ref{359})
is nonnegative and is identified as the entropy production rate
\cite{tome2010}
\begin{equation}
\Pi = \sum_i \frac{m^2}{\alpha_i T_i}\int \frac{J_i^2}{P} dxdv.
\label{339b}
\end{equation}
Although this identification may seem to be arbitrary, 
as has been argued \cite{spinney2012}, we will
see in the next section that in fact it is in accordance with 
the expression (\ref{9}).
It vanishes only when $J_i=0$ which is the equilibrium condition.
The second summation is thus the entropy flux
\begin{equation}
\Phi = -\sum_i \frac{m}{T_i}\int v_i J_i dxdv,
\label{339a}
\end{equation}
which can also be written as
\begin{equation}
\Phi = \sum_i \frac1{T_i}
\left(\alpha_i \langle v_i^2\rangle - \frac{\alpha_i T_i}{m}\right).
\end{equation}
After replacing $J_i$, given by (\ref{340}), into
(\ref{339a}) and performing an integration by parts,
the variation of the entropy of the system becomes
\begin{equation}
\frac{dS}{dt} = \Pi - \Phi.
\end{equation}

Let us assume that the forces are conservative,
$F_i=-\partial V/\partial x_i$. In this case,
we define the energy of the system as 
\begin{equation}
E(x,v) = \frac{m}2\sum_i v_i^2 + V(x).
\end{equation}
Using the Fokker-Planck equation in the form (\ref{348})
we get the following
expression for the time derivative of the average energy
$U=\langle E(x,v)\rangle$ 
\begin{equation}
\frac{dU}{dt} = -\Phi_u,
\end{equation}
where  
\begin{equation}
\Phi_u = -  \sum_i m \int v_i  J_i  dxdv
\end{equation}
is the flux of energy from the system to outside. 
To reach this expression we have performed appropriate
integration by parts and assumed that $P$ and its derivatives
vanish at the boundaries of integration. Using the definition
of $J_i$, given by (\ref{340}), we may write the energy flux as
\begin{equation}
\Phi_u =  \sum_i \alpha_i \left( \langle v_i^2 \rangle
- \frac{k_BT_i}{m}\right).
\end{equation}

When all temperatures are the same $T_i=T$
we have $\Phi=\Phi_u/T$ so that
\begin{equation}
\frac{dS}{dt} - \frac1T\frac{dU}{dt} = \Pi. 
\label{361}
\end{equation}
From which follows that the time variation of
$F=U-TS$ is given by $dF/dt=-T\Pi$ so that $dF/dt\leq0$.

Thermodynamic equilibrium occurs when all temperatures are
the same, $T_i=T$, and the forces are conservative,
$F_i=-\partial V/\partial x_i$. In this case, $K_i=0$ and
$J_i=0$, which leads to the following result for
the equilibrium probability distribution
\begin{equation} 
P^e(x,v) = \frac1Z e^{-E(x,v)/k_BT},
\end{equation}
where $Z$ is a normalization constant.

If we integrate equation (\ref{361}) in time, from an initial time
$t_0$ until infinity, when the system is in equilibrium, we get
\begin{equation}
S-S_0 - \frac1T(U-U_0) \geq 0.
\label{362}
\end{equation}

Let us suppose that the system is in contact with just one
heat reservoir at temperature $T$
and that is temperature is varying slowly so that
$dT/dt=\alpha$ is small. This is again the quasi-static
process that we have already discussed.
In this case, the quantity $J_i$
will be of the order $\alpha$ so that $\Pi$ will be
of the order $\alpha^2$. On the other hand, 
$\Phi$ remains at the linear order in $\alpha$ and we
may write from (\ref{361}) $dS/dt=(1/T)dU/dt$. It follows
that the entropy and energy cannot be arbitrary but
are connected by the relation $TdS=dU$ so that
a system performing a quasi-static process may be considered 
to be in equilibrium. From the result
(\ref{362}) we see that the curve that connect $U$ and $S$
has the  property of convexity. To perceive this it
suffices to imagine that at the initial time $t_0$ the
energy $U_0$ and entropy $S_0$ correspond to values
of equilibrium at a certain temperature $T_0$.

%--------------------------------------
\subsection{Nonequilibrium stationary state}

Let us take a look at the 
the energy variation per unit time, or power, 
${\cal P}_i$ associated to the $i$-th particle,
% REVER o termo DISSIPATED POWER
given by
\begin{equation}
{\cal P}_i =
\langle v_i F_i \rangle + \frac{m}2\frac{d}{dt}\langle v_i^2\rangle,
\end{equation}
where the first term is associated to the dissipation due
to the force $F_i$ acting on the particle and the second
the time variation of its kinetic energy. 
Using the Fokker-Planck equation it is straightforward
to shown that 
\begin{equation}
{\cal P}_i = \alpha_i \langle v_i^2\rangle - \frac{\alpha_i T_i}{m}.
\end{equation}
Therefore, 
\begin{equation}
\Phi_u = \sum_i {\cal P}_i,
\end{equation} 
and
\begin{equation}
\Phi = \sum_i \frac{{\cal P}_i}{T_i},
\label{371}
\end{equation}
and we recall that the summation is over $i$ such that
$\alpha_i\neq0$ and $T_i\neq0$.

Let us consider the contact of the system with two reservoirs
$A$ and $B$ at temperatures $T_1$ and $T_2$, respectively.
The heat flux from reservoir $A$ to the system is given by 
\begin{equation}
{\cal J} =  
\sum_{i\in A} {\cal P}_i,
\end{equation}
where the summation is over the particles that are in contact
with the reservoir $A$. A similar expression holds 
\begin{equation}
{\cal J}' =  
\sum_{i\in B} {\cal P}_i,
\end{equation}
for the
heat flux ${\cal J}'$ from reservoir $B$ to the system.
In the stationary state, $\Pi=\Phi$, and taking into account
the expression (\ref{371}) for $\Phi$, we get
\begin{equation}
\Pi = \frac{{\cal J}}{T_1} + \frac{{\cal J}'}{T_2}.
\end{equation}
But in the stationary state $\Phi_u={\cal J}+{\cal J}'=0$
so that
\begin{equation}
\Pi = X {\cal J},
\end{equation}
where
\begin{equation}
X = \frac1{T_1}-\frac1{T_2},
\end{equation}

Let us assume that $X$ is small so that 
$\Delta T=T_2-T_1$ is small. In this case 
${\cal J} = L \Delta T$ where $L$ is the thermal coefficient.
Writing the probability distribution as
$P(x,v)=P^e(x,v)[1 - \Delta T a(x,v)]$,
where $P^e(x,v)$ is the equilibrium distribution when
the temperatures of the reservoir is $T_1$, we 
may calculate ${\cal J}$ to get
\begin{equation}
L =  \sum_{i\in A} \alpha_i \int v_i^2 a(x,v) P^e(x,v)dxdv, 
\end{equation}
which may be undertood as an
average over the equilibrium distribution.

%----------------------------------------------------------
\section{Master equation representation of the Fokker-Planck equation}

%--------------------------------------
\subsection{Microcanonical ensemble}

It is possible to represent the Fokker-Planck in terms
of a master equation. This can be done by a discretization
of the phase space in a such a way that the continuum
limit will reduce the master equation to the Fokker-Planck
equation. From the representation we can easily identify
the transition probabilities from which we can obtain,
for instance, the entropy production rate.

To set up the discrete stochastic dynamics we imagine
a representative point in the phase space following
a stochastic trajectory. We consider two types of
transitions from a given point in the phase space.
The first type is defined by the transitions determined
by the Hamiltonian flow. This type of transition is defined by
\begin{equation}
(x,v)\to(H_i^+x,H_i^+v),
\label{402}
\end{equation}
where $H_i^+x$ and $H_i^+v$ are vectors with the same components of
the vectors $x$ and $v$ except the $i$-th components $x_i$
and $v_i$ which are transformed to $x_i'$ and $v_i'$, 
where $x_i'$ is given by
\begin{equation}
x_i'=x_i+bv_i,
\label{405}
\end{equation}
and $v_i'$ is determined in such a way that
the energy is conserved, that is,
\begin{equation}
E(H_i^+x,H_i^+v)=E(x,v),
\label{425}
\end{equation}
where $b>0$ is a parameter. Each transition occurs 
with rate $1/b$. If $b$ is sufficient small, $v_i'$
is given by
\begin{equation}
v_i' = v_i+b\frac{F_i}{m}.
\end{equation}
Notice that the Hamiltonian transition defined above
has no reverse in the sense that from a point 
$(H_i^+x,H_i^+v)$ we cannot reach the 
point $(x,v)$ with this type of transition.

The transitions of the second type changes only the
velocities and preserves the kinetic energy. This type
is defined by 
\begin{equation}
(x,v)\to(x,M_{ij}v),
\label{411}
\end{equation}
where $M_{ij}v$ is a vector with the same components
of the vector $v$ except the components $i$ and $j$
which are $v_i'$ and $v_j'$ given by
\begin{equation}
v_i' = v_i\cos\theta - v_j\sin\theta,
\qquad
v_j' = v_i\sin\theta + v_j\cos\theta,
\end{equation}
where $\theta>0$, 
so that $(v_i')^2+(v_j')^2=v_i^2+v_j^2$ and the
kinetic energy is preserved. Another possible transition
is defined by
\begin{equation}
(x,v)\to(x,M_{ji}v).
\end{equation}
Each of these transition
occurs with rate equal to $\lambda_{ij}/m^2\theta^2$.
Notice that this second type of transition has
a reverse since from the point $(x,M_{ij}v)$
it is possible to reach the point $(x,v)$. It
suffice to observe that $M_{ji}(M_{ij}v)=v$.

The transitions above lead us to the following master 
equation
\[
\frac{\partial}{\partial t}P(x,v) 
=  \sum_i \frac1b\{ P(H_i^-x,H_i^-v) - P(x,v) \}
\]
\begin{equation}
+ \sum_{ij} \frac{\lambda_{ij}}{2\varepsilon^2}
\{P(x,M_{ij}v) + P(x,M_{ji}v) - 2P(x,v) \},
\label{347}
\end{equation}
where $H_i^-$ is defined in a way similar do $H_i^+$
except that the sign in front of $b$ in equation (\ref{405})
is negative.
It is straightforward to show that expression (\ref{347}) 
reduces to the Fokker-Planck (\ref{349}) in the limit
$\varepsilon\to0$ and $b\to0$.

Taking into account that all transitions preserve the
energy $E(x,v)$ we see that in equilibrium the
probability distribution $P^e(x,v)$ depends on
$(x,v)$ through $E(x,v)$. If at initial time
the energy is equal to $U$, then $P^e(x,v)$ 
vanishes if $E(x,v)\neq U$ and is a constant
if $E(x,v)=U$, which is the Gibbs microcanonical distribution.

%--------------------------------------
\subsection{Canonical ensemble}

Next we set up a discrete stochastic dynamics, described
by a master equation, whose continuous limit gives the
Fokker-Planck equation (\ref{315}). The representative
point in phase space $(x,v)$ performs a stochastic trajectory.
We consider again two types of
transitions from a given point in the phase space.
The first type is the transition defined by the Hamiltonian
flow given by equation (\ref{402}).
The transition of the second type changes only the
velocities but in general it does not preserve the kinetic energy.
This type of transition is defined by 
\begin{equation}
(x,v)\to(x,C_i^\pm v),
\label{419}
\end{equation}
where $C_i^\pm v$ is a vector with the same components of
the vector $v$ except the $i$-th component $v_i'$ which
is given by
\begin{equation}
v_i' = v_i\pm a,
\label{321a}
\end{equation}
where $a>0$ is a parameter and each transition
occurs with rate 
\begin{equation}
A_i^{\pm}(v) = \frac{\alpha_i k_BT_i}{m^2a^2}e^{\mp amv_i/2k_BT_i}.
\label{328a}
\end{equation}

The master equation is written as 
\[
\frac{\partial}{\partial t}P(x,v) =
\sum_i \frac1{b}\{ P(H_i^-x,H_i^-v) - P(x,v) \}
\]
\[
=\sum_i \{ A_i^+(C_i^-v)P(x,C_i^-v) - A_i^-(v)P(x,v) \}
\]
\begin{equation}
+\sum_i \{ A_i^-(C_i^+v)P(x,C_i^+v) - A_i^+(v)P(x,v) \}.
\end{equation}

It is straighfoward to show that in the limit $a\to0$ and $b\to0$,
the master equation reduces to equation (\ref{315}),
and the master equation can indeed be understood as a 
representation of the Fokker-Planck equation (\ref{315}).

The stationary solution of the master equation
when all temperatures are the same, which corresponds
to the thermodynamic equilibrium, is the Gibbs distribution.
Indeed, the detailed balance of the master equation gives us
the relation
\begin{equation}
P^e(H_i^-x,H_i^-v) = P^e(x,v),
\end{equation}
which means that $P^e(x,v)$ depends on $(x,v)$ through $E(x,v)$.
Writing 
\begin{equation}
P(x,v) = \frac1Z e^{- E(x,v)/k_BT},
\label{322}
\end{equation}
we see that the other relation,
\begin{equation}
A_i^+(C_i^-v)P^e(x,C_i^-v) = A_i^-(v)P^e(x,v),
\end{equation}
is fullfiled if we take into account that all 
temperatures are the same, $T_i=T$.

%--------------------------------------
\subsection{Entropy production}

We have seen that the entropy production rate
of a system described by a master equation is
obtained by expression (\ref{9}). This expression
is appropriate when the rates of the reversed transitions
are nonzero. This is the case of transitions defined
by (\ref{411}). The entropy production rate $\Pi_M$ associated
to these transitions, according to (\ref{9}) is given by
\[
\Pi_M = \frac{k_B}2\sum_{x,v} \sum_{ij}\frac{\lambda_{ij}}{2m^2\theta^2}
\{P(x,M_{ij}v)-P(x,v)\}\ \times
\]
\begin{equation}
\times \ln\frac{P(x,M_{ij}v)}{P(x,v)}.
\label{423}
\end{equation}
In the limit $\theta\to0$, the right-hand side reduces to
expression on the right-hand side of (\ref{413}).
The entropy flux $\Phi_M$ associated to the transitions (\ref{411})
is obtained by using (\ref{9fl}), but 
it vanishes identically,
\begin{equation}
\Phi_M = 0.
\label{423a}
\end{equation}

Let us consider now the entropy production rate $\Pi_C$
associated to the transitions defined by (\ref{419}).
According to expression (\ref{9}), it is given by
\[
\Pi_C =\frac{k_B}2
\sum_{x,v}\sum_i\{A_i^+(C_i^-v)P(x,C_i^-v)-A_i^-(v)P(x,v)\}\times
\]
\begin{equation}
\times\ln\frac{A_i^+(C_i^-v)P(x,C_i^-v)}{A_i^-(v)P(x,v)}.
\label{424}
\end{equation}
After taking the limit $a\to0$ this expression is reduced to
the result (\ref{339b}). The corresponding entropy flux 
$\Phi_C$ is obtained from (\ref{9fl}) and is given by
\[
\Phi_C =\frac{k_B}2
\sum_{x,v}\sum_i\{A_i^+(C_i^-v)P(x,C_i^-v)-A_i^-(v)P(x,v)\}\times
\]
\begin{equation}
\times\ln\frac{A_i^+(C_i^-v)}{A_i^-(v)}.
\end{equation}
The limit $a\to0$ leads us to the result (\ref{339a}).

We now wish to consider the entropy production rate
and the flux of entropy 
coming from the parts of the stochastic trajectory
associated to the Hamiltonian flow, given by the
transitions defined by (\ref{425}).
We postulate that the entropy flux associated 
to the Hamiltonion flow vanishes identically, $\Phi_H=0$.
Therefore, the entropy production rate associated to the
Hamiltonian flow should be equal to part of $dS/dt$ coming from 
the Hamiltonian flow. This part can be obtained by
inserting the first summation of the right-hand side of
the master equation (\ref{347}) into equation (\ref{26a}). After
doing this, we get the following expression for the
entropy production rate associated to the Hamiltonian flow,
\begin{equation}
\Pi_H = k_B\sum_{x,v} \sum_i\frac1{b}
P(x,v)\ln \frac{P(x,v)}{P(H^+x,H^+v)},
\label{445a}
\end{equation}
which is similar to expression (\ref{9r}).

Next we have to show that $\Pi_H\geq0$. To this end,
we expand each term in the summation in powers of $b$.
Up to linear terms in $b$ the $i$ element of the summation
equals
\begin{equation}
\left(v_i\frac{\partial P}{\partial x_i}
+\frac{F_i}{m}\frac{\partial P}{\partial v_i}\right)
+ \frac{b}{2P}\left(v_i\frac{\partial P}{\partial x_i}
+\frac{F_i}{m}\frac{\partial P}{\partial v_i}\right)^2.
\end{equation}
But the integral in $x$ and $v$
of the first term vanishes so that $\Pi_H\geq0$.
In fact, it vanishes in the
continuum limit $b\to0$.
Therefore, in the continuum limit $\Pi_H=0$.
From this result it follows that $\Pi_M$ is the
total production of entropy for the microcanonical case.
Since in the continuous limit, it goes into
(\ref{350a}), it follows that the expression given 
by (\ref{350a}) is indeed the entropy production
rate, as we have assumed. Similarly, it follows that
$\Pi_C$ is the total production rate for the canonical case.
In the continuous limit it is identified with (\ref{339b})
so that the expression given by (\ref{339b}) is indeed
the entropy production rate, as assumed.

%----------------------------------------------------------
\section{Conclusion}

We developed the stochastic approach to thermodynamics based
on the stochastic dynamics. More specifically, we used 
the master equation, in the case of discrete state space, and
the Fokker-Planck, in the case of continuous state space.
Our approach is founded on the use of a form for the production 
of entropy which is nonnegative by definition and vanish
in equilibrium. 
Based on these assumptions we studied interacting systems
with many degrees of freedom in equilibrium or out
of thermodynamic equilibrium and how the macroscopic laws can be
derived from the stochastic dynamics.
This required the introduction of the transition rates
which thus play a fundamental role in the present approach,
similar to the Gibbs distribution in the case of equilibrium.

Using the property that the production of entropy is
nonnegative, which is understood as the dynamic formulation
of the second law of thermodynamics, we were able to 
show that in the quasi-static process, the representative point in
the thermodynamic space approaches a surface 
and that this surface has the property of convexity.
These statements are usually introduced as postulates in
equilibrium thermodynamics.
We have also shown the bilinear form of entropy
production, which is a sum of terms, each one being a product
of a force and a flux. We remark that this is the macroscopic
form used in nonequilibrium thermodynamics, 
and should not be confused with
the microscopic definition of entropy production itself, which
looks like a bilinear form.
From the bilinear form of the entropy production, we have
determined the Onsager coefficients and shown that they
obey the reciprocal relations. 
The nonequilibrium steady states of a system with
several chemical species and chemical reactions
were studied by the use of appropriate transition rates.
From the definition of the entropy production rate
it was possible to derive the bilinear form, which
in this case is written in terms of affinities and
the rates of the extents of reaction. In equilibrium
the affinities vanish, which is the condition for
chemical equilibrium. 

Using appropriate transition rates or appropriate stochastic 
noise, in the case of the Fokker-Planck, it was possible to
study several situations that were analogous to
those related to the microcanonical, canonical and
grand canonical Gibbs ensembles. For the microcanonical
case in continuous state space we have introduced an
energy conserving stochastic noise. For the canonical
case we used the usual white Gaussian noise.
To make contact with the master equation, we have
used a master equation representation of the Fokker-Planck.
Using this representation we confirmed the expression for the
production of entropy that was introduced by the
splitting of the time derivative of entropy.
In this case we postulated that a Hamiltonian transition
induces no flux of entropy. Since the entropy is constant
along a Hamiltonian flow in continuous space, this implies
no production of entropy.

%-------------------------------------------------------

\end{document}